\documentclass{elsarticle}
\usepackage[utf8]{inputenc}
\usepackage[a4paper, total={6in, 10in},]{geometry}
\usepackage{amsmath}
\usepackage{amsfonts}
\usepackage{bbold}
\usepackage{amssymb}
\usepackage{amsthm}
\usepackage{mathtools}

\usepackage{mathdots}
\usepackage[table]{xcolor}

\usepackage{hyperref}
\hypersetup{
	colorlinks=true,     
	citecolor=green!60!black,      
	linkcolor=black,      
}

\usepackage{graphicx}
\usepackage{tikz}
\usepackage[ruled,linesnumbered]{algorithm2e}

\SetKwInOut{Input}{Input}
\SetKwInOut{Output}{Output}
\SetKwInOut{Oracle}{Oracle}

\newtheorem{problem}{Problem}
\newtheorem{theorem}{Theorem}

\newtheorem{proposition}{Proposition}

\newtheorem{lemma}{Lemma}
\newtheorem{corollary}{Corollary}

\makeatletter
\DeclareRobustCommand\onedot{\futurelet\@let@token\@onedot}
\def\@onedot{\ifx\@let@token.\else.\null\fi\xspace}
\def\ie{{\it i.e\onedot, }} 
\makeatother

\renewcommand{\k}{\kappa}
\renewcommand{\bar}{\overline}
\renewcommand{\tilde}{\widetilde}
\newcommand{\mbf}{\mathbf}

\DeclareMathOperator{\rk}{rk}

\DeclareMathOperator{\herm}{herm}

\DeclareMathOperator{\res}{Res}
\DeclareMathOperator{\spn}{span}
\DeclareMathOperator{\LCLM}{LCLM}

\DeclareMathOperator{\diag}{Diag}
\DeclareMathOperator{\Deg}{Deg}
\DeclareMathOperator{\degMM}{deg_{M}}
\DeclareMathOperator{\Minordeg}{Minor-deg}

\newcommand{\DP}{\partial}

\title{Degree bounds for linear differential equations and recurrences}
\author[1]{Louis Gaillard\corref{cor1}%
}

\affiliation[1]{organization={ENS de Lyon, CNRS, Inria,
    Université Claude Bernard Lyon 1, LIP, UMR 5668 },
city={Lyon},
country={France}}

\cortext[cor1]{Corresponding author: louis.gaillard@ens-lyon.fr}

\begin{document}

\begin{abstract} 
  Linear differential equations and recurrences reveal many properties
  about their solutions. Therefore, these equations are well-suited for
  representing solutions and computing with special functions.  We
  identify a large class of existing algorithms that compute such
  representations as a linear relation between the iterates of an
  elementary operator known as a \emph{pseudo-linear map}.  Algorithms of
  this form have been designed and used for solving various computational
  problems, in different contexts, including effective closure
  properties for linear differential or recurrence equations, the
  computation of a differential equation satisfied by an algebraic
  function, and many others.
  We propose a unified approach for establishing precise degree bounds
  on the solutions of all these problems. This approach relies on a
  common structure shared by all the specific instances of the class.
  For each problem, the obtained bound is tight.
  It either improves or recovers
  the previous best known bound that was derived by \emph{ad hoc}
  methods.
\end{abstract}


\maketitle

\section{Introduction}
\label{sec:introduction}

Numerous elementary and special functions (resp.~sequences) may be
defined as a solution of a linear differential (resp.~recurrence)
equation with polynomial coefficients.  Equivalently, one can view
such an equation as a linear differential (resp.~recurrence) operator
that acts on functions and annihilates solutions.  The solutions to
these equations define the class of D-finite functions
(resp.~P-recursive sequences)~\cite{stanley1980dfinite,Kauers23}.  Many
mathematical properties satisfied by D-finite functions can be
directly computed from their defining equation.
This motivates a fruitful line of research in computer algebra, that
consists in considering operators as an efficient data structure for
representing their solutions~\cite{salvy2019survey,salvy1994gfun}.
This paradigm requires efficient routines for computing with this
data structure and implementing the \emph{arithmetic of operators}.

Let $k$ be a field of characteristic 0 and $x$ an indeterminate.  We
denote by $\DP_x$ the usual derivation~$d/dx$ on the rational function
field $k(x)$ and by $\sigma_x$ the shift endomorphism $x \mapsto x+1$,
that maps any rational function $f(x)$ to $f(x+1)$.  By a slight abuse
of notation, we denote by $k(x)\left<\DP\right>$ where $\DP$ is a
derivation on $k(x)$, the skew polynomial ring of linear ordinary
differential operators in $\DP$, and $k(x)\left<\sigma_x\right>$ for
the skew polynomial ring of linear ordinary recurrence
operators. Those are two examples of Ore
algebras~\cite{Ore1933,BronPetk96}.  Up to multiplying by the
denominator, one can always manipulate operators with polynomial
coefficients in $k[x]$.  In this work, we study three families of
computational problems for operators.

\paragraph*{Closure operations}
D-finite functions are closed under sums and
products~\cite[Thm.~2.3]{stanley1980dfinite}.  For example, if two
functions $\alpha_1$ and $\alpha_2$ are respectively represented by an
annihilating operator $L_1$ and~$L_2$, one wants to compute a
representation for $\alpha = \alpha_1 + \alpha_2$, that is, a nonzero
operator $L$ such that~$L(\alpha) = 0$. Thus, the effective closure by
sum is the following problem: given
$L_1,L_2 \in k[x]\left<\DP_x\right>$, compute
$L \in k[x]\left<\DP_x\right> \setminus \{0\}$ that annihilates all
the sums $\alpha_1 + \alpha_2$, where
$L_1(\alpha_1) = L_2(\alpha_2) = 0$.  If one also requires $L$ to have
minimal order, then it is a \emph{least common left multiple (LCLM)}
of~$L_1$ and $L_2$. Several algorithms have been designed and analysed
for computing LCLMs~\cite{BoChSaLi2012,vdHoeven2016}.  For the closure
by product, the analogous notion is called a \emph{symmetric product}.
More generally, one can handle closure of D-finite functions under any
polynomial operation of the following type: given operators~$L_1, \dots,
L_s \in k[x]\left<\DP_x\right>$, compute a nonzero linear
differential operator $L$ such that for arbitrary solutions $\alpha_i$
of $L_i$ for all $1 \le i \le s$, $L$ annihilates
$\alpha = J(x,\alpha_1,\dots, \alpha_1^{(i_1)},\dots,\alpha_s, \dots,
\alpha_s^{(i_s)})$ where $J$ is a fixed multivariate polynomial.
Algorithmic approaches also exist at this level of
generality~\cite{Kauers2014}.  All these problems directly translate
in the shift case for recurrence operators.

\paragraph*{Algebraic power series}
A power series $\alpha(x)$ over the field $k$ is said to be
\emph{algebraic} if it is a root of a nonzero bivariate
polynomial~$P(x,y) \in k[x,y]$.
It is known that such an algebraic function is
also D-finite~\cite[p.~287]{abeloeuvres}, and we call
\emph{differential resolvent} an operator annihilating all the roots
of~$P$.  Cockle provided an algorithm~\cite{cockle1861} for computing
differential resolvents, that is the basis of a more recent efficient
algorithm for this problem~\cite{BoChSaLeSc2007}.  The method extends
to the problem of the computation of an annihilating operator for the
composition~$h = f \circ g$ of a D-finite power series~$f$ and an
algebraic power series $g$~\cite{KaPo2017}, that is also known to be
D-finite~\cite[Thm.~2.7]{stanley1980dfinite}.

\paragraph*{Creative telescoping for bivariate rational functions}
Creative telescoping refers to a family of methods that have proved
successful for symbolic integration and summation.  The general
concept was first introduced by Zeilberger~\cite{zeilberger1990} and
it applies to a large class of functions.  Here we focus on the simple
integration of a bivariate rational function
$f(x,y) = p /q \in k(x,y) \setminus \{ 0 \}$ with $p$ and $q$ coprime.
For any integration contour $\omega$ that avoids the poles of $f$, the
definite integral $F(x) = \int_\omega f(x,y) \, dy$ is known to be
D-finite and the problem consists in computing a linear differential
equation satisfied by~$F$.  The heart of the method relies on finding
a \emph{telescoper} for~$f$, namely a linear differential operator
$L \in k[x]\left<\DP_x\right> \setminus \{ 0 \}$ such that
$L( f(x,y)) = \DP_y ( h(x,y))$ for some rational
function~$h$. Under reasonable assumptions on~$h$ and~$\omega$, the
operator $L$ is then proved to annihilate $F$. We specifically study
an algorithm~\cite{BoChChLi2010} based on successive Hermite
reductions to address the problem of computing a telescoper for
$f(x,y)$.

\vspace{5pt}

Despite the diversity of these problems, they all admit algorithmic
solutions where the coefficients of the computed operator are obtained
as an element in the nullspace of a matrix of rational functions, that
has a certain \emph{structure}.  The common principle in these
algorithms, that explains the structure in the matrix, is the
following: consider an arbitrary solution $\alpha(x)$ of the computed
operator $L$; this solution $\alpha$ is defined implicitly by a set of
relations it satisfies. These relations are given by the definition of
$L$ and allow to see $\alpha$ as an element $a_0$ of a finite
dimensional vector space $\mathcal A$ over $k(x)$.  Then, using the
relations satisfied by $\alpha$, each successive derivative
$\alpha^{(i)}$ is also identified to an element~$a_i$ of $\mathcal A$,
and finally, the operator $L$ is obtained as a linear relation between
the vectors~$(a_i)_{i \ge 0}$. Such a relation exists by the finite
dimension of $\mathcal A$.

For example, for computing an LCLM of
$L_1$ and $L_2$, consider an arbitrary solution~$\alpha =
\alpha_1 + \alpha_2$ defined by the relations
$L_1(\alpha_1) = L_2(\alpha_2) = 0$. Then, express the successive
derivatives of $\alpha$ using~$L_1$ and $L_2$ to replace high order
derivatives of $\alpha_1$ and $\alpha_2$, until a linear relation is
found.  Cockle's algorithm~\cite[Sec.~2.1]{BoChSaLeSc2007} computes a
differential resolvent of $P \in k[x,y]$ using an analogous
strategy:~$\alpha$ is defined by $P(x,\alpha) = 0$, then, the successive
derivatives of $\alpha$ can be seen as elements of the quotient ring
$k(x)[y]/(P)$, that is a finite dimensional vector space over $k(x)$.

This general principle can be formalised by the following algebraic
setup.  We are given a sequence~$(a_i)_{i\ge 0}$ of $k(x)^n$ defined
by the simple inductive formula:
\begin{align}
  \label{eq:def_ai}
  a_0 = a \in k[x]^n, \quad a_{i+1} = \theta (a_i),~ i \ge 0,
\end{align}
where $\theta : k(x)^n \rightarrow k(x)^n$ is a \emph{pseudo-linear
  map}~\cite{Jacobson37, BronPetk96}, that is, there exists a linear
map $T$, seen as a matrix in $k(x)^{n \times n}$, such that
$\theta = \DP + T$ in the differential case, or
$\theta = T \cdot \sigma_x$ in the shift case.  One may interpret the
matrix $T$ as carrying out the action of the derivation (resp. the
shift) modulo some relations.  The problem is to compute a minimal
linear relation with polynomial coefficients between the first
elements of the sequence $(a_i)_{i \ge 0}$.
\begin{problem}
  \label{pb:lin_rel_pseudo_lin_map}
  Given $\theta$ a pseudo-linear map on $k(x)^n$, $a \in k[x]^n$ and
  letting $\rho = \dim\spn_{k(x)}(\theta^i a, i \ge 0) \le n$, find
  $\eta = (\eta_0,\dots, \eta_{\rho}) \in k[x]^{\rho+1} \setminus \{ 0
  \}$ such that
  \begin{equation}
    \label{eq:lin_rel_pseudo_lin_map}
    \eta_0 \cdot a +  \eta_1 \cdot \theta a + \cdots + \eta_\rho \cdot \theta^\rho a = 0.
  \end{equation}
\end{problem}
In the case $\rho = n$, the vector $a$ is called a cyclic vector of
$\theta$~\cite{adjamagbo1988, Barkatou93,BoChPa2013,churchill2002} and
the above relation gives a characteristic polynomial of
$\theta$~\cite{amitsur1954}.  By analogy with the classical Krylov
method for computing the characteristic polynomial of a matrix
$T$~\cite{keller-gehrig1985, NePeVi2024}, the linear
system~(\ref{eq:lin_rel_pseudo_lin_map}) is referred to as a
\emph{pseudo-Krylov} system.
Each of the problems mentioned above reduces to solving the pseudo-Krylov
system~(\ref{eq:lin_rel_pseudo_lin_map}) for a specific choice of
$\theta$ (and thus of the linear map $T$) and $a$. 
Moreover, each reduction
to Problem~\ref{pb:lin_rel_pseudo_lin_map} corresponds to an existing
algorithm in the literature~\cite{stanley1980dfinite,Kauers2014,
  BoChSaLeSc2007,KaPo2017,BoChChLi2010}.  The corresponding algorithm
constructs and solves the associated pseudo-Krylov system.  Each
algorithm has usually the best known complexity for solving the
problem (except for LCLMs, where there exist faster algorithms using
Sylvester-type matrices adapted to the skew
setting~\cite{BoChSaLi2012}, or an evaluation-interpolation approach
based on truncated fundamental basis of
solutions~\cite{vdHoeven2016}).  However, for all specific instances,
the degree bound on the solution $\eta$ that is derived from the
linear system~(\ref{eq:lin_rel_pseudo_lin_map}) is an overestimation
of the actual degree.  Tighter bounds are sometimes established and
obtained by \emph{ad hoc}
constructions~\cite{BoChSaLi2012,BoChSaLeSc2007,BoChChLi2010}, or only
conjectured~\cite{bostan2003these,KaPo2017}.
This degree gap suggests the existence of an unrevealed structure
in the pseudo-Krylov system relative to each instance. 
We aim at understanding this structure to accelerate the resolution
of the system~(\ref{eq:lin_rel_pseudo_lin_map}) for all specific instances.

Our main contribution is to provide a refined technique for proving
degree bounds, and thus to explain the degree gap.
In each instance, a tight bound is obtained by viewing the matrix
of rational functions $T$
as performing reduction modulo some relations. Indeed, we show that
this enables to find a certain \emph{realisation} of $T$~\cite{coppel1974},
namely, writing
\begin{align*}
  T = XM^{-1}Y,
\end{align*}
where $X,M,Y$ are polynomial matrices of \emph{small} degree of sizes
$n \times h, h \times h, h\times n$ respectively. Since
the matrices $X,M,Y$ have small degrees compared to the matrix $T$,
such a realisation yields a much more compact representation of $T$, providing
tight degree bounds and explaining the gap.
This is an
extension
to more general pseudo-Krylov systems
of the recent approach~\cite{gaillard24}
where only the differential
case $\theta = \DP_x + T$ is handled.
Also, the previous approach requires the matrix $T$ to be
\emph{strictly proper} \textemdash \ namely, each entry of $T$ has
numerator with degree strictly less than the denominator~\textemdash \
as a technical assumption for the proof. Due to this assumption, the
degree bound obtained in each problem from this approach is only
valid under a genericity or a regularity assumption on the input.  The
extension we now present is freed from the strict properness
assumption, leading to a unified approach for proving
bounds that hold unconditionally in each
instance.

\subsection{Overview of the method}
\label{sec:overview_method}

Our approach for establishing bounds mostly relies on tools
originating from control theory for linear systems over $k(x)$.  For a
matrix of rational functions $R \in k(x)^{n \times p}$, one can define
an appropriate notion of degree, the \emph{McMillan degree}
of $R$~\cite{mcmillan1952I,mcmillan1952II}, denoted as
$\degMM(R)$. The McMillan degree extends the definition of degree of a
rational function as its number of poles counted with multiplicity,
including those at infinity.  We relate $\degMM(R)$ with the minimal
degree of an element of the (right) nullspace of $R$.  Actually, if
$\rho$ denotes the rank of $R$ and $\rho < p$, we show that there
exists a non-trivial polynomial solution to the linear system
$R \eta = 0$ with
\begin{align*}
  \deg(\eta) \le \frac{\degMM(R)}{p-\rho}.
\end{align*}
The precise definitions and results can be found in
Section~\ref{sec:mat_rat_function}~(see~Theorem~\ref{thm:McMillandeg_kronecker}).
This bound follows from classical results on minimal bases of
$k[x]$-modules~\cite{forney1975}. It can also be seen as an extension
to matrices of rational functions of the degree transfer between a
polynomial matrix and a minimal nullspace
basis~\cite[Thm.~3.3]{StoVi2005issac}~(see
also~\cite{StoVi2005report,BeLaVi2006} for a proof).

Next, our degree bounds for solutions of
Problem~\ref{pb:lin_rel_pseudo_lin_map} follow from this result
applied to the matrices defining the pseudo-Krylov
systems~(\ref{eq:lin_rel_pseudo_lin_map}); the difficulty is then to bound
their McMillan degree.  Those matrices are called
\emph{pseudo-Krylov matrices} and are of the form $K =
\begin{bmatrix}
  a & \theta a & \cdots & \theta^m a 
\end{bmatrix} \in k(x)^{n \times (m+1)}$, where $a \in k[x]^n$ is a vector of polynomials
of degree at most $d_a$ and $\theta$ a pseudo-linear map
of the following shapes. There exists a matrix $T \in k(x)^{n \times n}$ such that
$\theta =
p(x) \DP_x + T$ where~$p$ is a polynomial of degree at most $1$ in the differential case,
or $\theta = T \sigma_x$ in the shift case. These two cases cover a class of Ore
operators containing differential operators in the usual derivation $\DP_x$ or in
the Euler derivation $x\DP_x$ and classical recurrence operators in $\sigma_x$.
In the differential case, the assumption on the degree of $p$ ensures that
the derivation $\DP = p\DP_x$ does not add any pole (including infinity)
when differentiating a rational function.
In Section~\ref{sec:deg_bounds_pseudo_krylov}, we study and bound the McMillan
degree of this class of pseudo-Krylov matrices. 
We show in Theorem~\ref{thm:bd_mcmillan_pseudoKrylov} that
\begin{align*}
  \degMM(K) \le \rho d_a + m \degMM(T).
\end{align*}
As a consequence, we establish our main result.
\begin{theorem}
  \label{thm:degree_bound_pseudo_krylov}
  Let $\theta$ be a pseudo-linear map of the shape defined above with
  a matrix $T \in k(x)^{n \times n}$~($\theta = p(x)\DP_x + T$ with
  $p \in k[x]$ of degree $\le 1$, or $\theta = T \cdot \sigma_x$).
  Let $a \in k[x]^n$ of degree $d_a$ and
  $\rho = \dim \spn_{k(x)}(\theta^i a, i \ge 0)$.  Then, for all
  $m \ge \rho$ there exists a linear relation
  $ \eta_0 \cdot a + \cdots + \eta_m \cdot \theta^m a = 0, $ with
  $\eta = (\eta_0,\dots,\eta_m) \in k[x]^{m+1}\setminus \{0\}$ such
  that
  \begin{align*}
    \deg(\eta) \le 
    \left(\rho d_a + m \degMM(T)\right) / (m+1-\rho).
  \end{align*}
  In particular, there exists a solution
  $\eta \in k[x]^{\rho+1}\setminus \{0\}$
  of Problem~\ref{pb:lin_rel_pseudo_lin_map} for $(\theta,a)$
  satisfying
  \begin{align*}
    \deg (\eta) \le \rho d_a + \rho \degMM(T).
  \end{align*}
\end{theorem}

The coefficients of a solution of
Problem~\ref{pb:lin_rel_pseudo_lin_map} are thus bounded by
$O(nd_a + n\degMM(T))$. Besides, if one relaxes the minimality on the
order of the operator in output in
Problem~\ref{pb:lin_rel_pseudo_lin_map} and allows for relations of
length $m+1 \ge \rho+1$, we give a degree bound on a solution of
minimal degree of the over-determined pseudo-Krylov system that is
parameterised by $m$.  When applied to a specific instance, this gives
degree bound for operators that is parameterised by the order.  The
obtained parameterisations are called \emph{order-degree curves} and
explain the observation that higher order operators generally have
lower degree~\cite{ChJaKaSi2013,ChKaSi2016}.

By Theorem~\ref{thm:degree_bound_pseudo_krylov}, we are led to
estimate the McMillan degree of the matrix $T$ of the problem in order
to establish tight degree bounds.  In
Sections~\ref{sec:lclm}-\ref{sec:hermite_red_bd}, we bound $\degMM(T)$
for each problem mentioned before,
using a realisation $XM^{-1}Y$ for $T$.
Indeed, a tight bound on $\degMM(T)$ can be derived from the degrees
of the polynomial matrices $X,M,Y$.

\subsection{Results in specific instances}
\label{sec:results_specific_instances}

Our unified approach can thus be summarised in two steps: find a
realisation of the matrix of the problem $T$ to bound $\degMM(T)$,
then apply Theorem~\ref{thm:degree_bound_pseudo_krylov}. The approach
is exemplified in Sections~\ref{sec:lclm}-\ref{sec:hermite_red_bd} for
all the problems mentioned before.  For each problem, a degree bound
is obtained for minimal order operators and an order-degree curve is
also
given~(Theorems~\ref{thm:degree_bound_LCLM}-\ref{thm:herm_red_bd}).
These results are compared with the best known bounds in the
literature. A summary is given in
Table~\ref{tab:results_bound_instances} for minimal order operators.
The bounds in the three first rows correspond to results of
Sections~\ref{sec:lclm}-\ref{sec:closure_polynomials} and deal with
closure properties of linear differential operators in
$\DP = p(x)\DP_x$ with $p \in k[x]$ of degree at most~$1$ and recurrence
operators in $\sigma_x$. For LCLMs and symmetric products of $s$
operators of order at most $r$ and degree at most $d$~(rows
\textsf{LCLM} and \textsf{SymProd}) the bounds are expressed in terms
of the parameters~$s, r, d$. The row \textsf{Polynomials} stands
for closure of $s$ operators under a polynomial operation defined by
$J$ (see Theorem~\ref{thm:poltodiffeq_bound} for details).  The two
next rows stand for the family of problems dealing with algebraic
functions and record results of
Sections~\ref{sec:algeqtodiffeq}-\ref{sec:subs_algDfinite}.  For a
differential resolvent of $P(x,y)$~(row \textsf{AlgeqtoDiffeq}), the
bounds are given in terms of $(d,r)$ the bi-degree of~$P$.  The row
\textsf{Composition} stands for the computation of an annihilator of
every function $h = f \circ g$ where~$f$ annihilates a fixed operator
$L \in k[x]\left<\DP_x\right>$ of degree $d$ and order $r$ and $g$ is
a root of a fixed bivariate polynomial $P(x,y)$ also of bi-degree
$(d,r)$. Finally the last row \textsf{Hermite} bounds the degree of a
minimal order telescoper of a rational function
$f(x,y) = p(x,y)/q(x,y)$ such that $(d,r)$ is the bi-degree of
$q(x,y)$ and $\deg_x(p) \le d$ and $\deg_y(p) < r$. The details can be
found in Section~\ref{sec:hermite_red_bd}.

\begin{table}[h]
  \centering \renewcommand{\arraystretch}{1.2}
  \begin{tabular}{c|c|c|c||cc}
    & $\rho$ & $\degMM(T)$ & New Bound & \multicolumn{2}{c}{Previous Bound}\\
    \hline
    \textsf{LCLM} & $sr$ & $sd$ &$s^2rd + o(s^2rd)$&
                                                     $s^2rd + o(s^2rd)$ &\cite{BoChSaLi2012}  \\
    \hline
    \textsf{SymProd} &$r^s$& $sr^{s-1}d$ &$sr^{2s-1}d $&
                                                         $sr^{2s}d$ &\cite{Kauers2014}\\
    \hline
    \textsf{Polynomials} &$R$
             & $R\sum_{i=1}^s \frac{k_i d_i}{k_i +r_i -1}$ &$ R (\deg_x(J) +
                                                             \degMM(T))$& $ R (\deg_x(J) +R\sum_{i=1}^sk_i d_i) $
                  &\cite{Kauers2014}\\
    \hline
    \textsf{AlgeqtoDiffeq} &$r$ & $(2r-1)d$  &$2r^2d + o(r^2d)$
                                       &$4r^2d + o(r^2d)$ &\cite{BoChSaLeSc2007} \\
    \hline
    \textsf{Composition} &$r^2$ & $(r(2r-1)+d)d$  &$O(r^2d^2 + r^4d)$
                                       &$O(r^3d^2 + r^4d)$ &\cite{KaPo2017} \\
    \hline
    \textsf{Hermite} &$r$ & $2rd$ &  $2r^2d + o(r^2d)$
                                       &$2r^2d + o(r^2d)$ &\cite{BoChChLi2010}
  \end{tabular}
  \caption{Degree bounds for instances of
    Problem~\ref{pb:lin_rel_pseudo_lin_map}}
  \label{tab:results_bound_instances}
  \renewcommand{\arraystretch}{1}
\end{table}

\section{Matrices of rational functions}
\label{sec:mat_rat_function}

We recall the needed results and definitions from the theory of
matrices of rational functions. In particular, we underline how the
structure of the minors of a matrix of rational functions, that is
captured by its so-called \emph{determinantal
  denominators}~\cite{coppel1974},
is connected to the size of a
minimal basis of the nullspace~\cite{forney1975}.  Most material we
need here originates from control theory (see
e.g.,~\cite{kailath1980}).

For a rational matrix $R \in k(x)^{n\times p}$ of rank $r$, the
(right) nullspace of $R$ is the $k(x)$-vector space of dimension $p-r$
containing all vectors $\eta \in k(x)^p$ such that $R\eta = 0$. One
can restrict to polynomial vectors $\eta \in k[x]^p$ and see the
nullspace as a $k[x]$-module~\cite{forney1975}.  For a polynomial
matrix~$M$ with~$c$ columns, the column degree of $M$ designates the
non-decreasing sequence~$v_1 \le \cdots \le v_c$ of the degrees of the
columns of $M$.  A basis $N \in k[x]^{p \times (p-r)}$ of the
nullspace is called a \emph{minimal basis} if the columns of $N$ have
non-decreasing degrees and the column degree
$v_1 \le \cdots \le v_{p-r}$ of $N$ is minimal among all nullspace
bases~$N'$, that is, the column degree $v_1' \le \cdots \le v_{p-r}'$ of $N'$
satisfies~$v_i' \ge v_i$ for all~$1\le i \le p-r$.  The
degrees~$v_1, \dots, v_{p-r}$ are structural invariants of the matrix
$R$ and are called the \emph{(right) Kronecker indices} of
$R$~\cite[Sec.~6.5.4, p.~455]{kailath1980}.  Our main goal is to bound
the minimal degree of a solution of pseudo-Krylov
systems~(\ref{eq:lin_rel_pseudo_lin_map}).  In other words, we want to
bound the first Kronecker index~$v_1$ of the associated pseudo-Krylov matrix.

The main result of this section is to relate the Kronecker
indices of a matrix of rational functions with its
McMillan degree. The definition of McMillan degree can be found in
Section~\ref{sec:mcmillan-degree}. 
The following
result extends to rational matrices the degree transfer between a polynomial
matrix and a minimal nullspace basis~\cite[Thm.~3.3]{StoVi2005issac}.

\begin{theorem}
  \label{thm:McMillandeg_kronecker}
  Let $R \in k(x)^{n \times p}$ of rank $r<p$ and Kronecker indices
  $v_1,\dots,v_{p-r}$. Then,
  \begin{align}
    \label{eq:kronecker_mcmillan}
    \sum_{i=1}^{p-r} v_i \le \degMM(R), \text{ and } v_1 \le \frac {\degMM(R)}{p-r} .
  \end{align}
\end{theorem}

It is sufficient to estimate the McMillan degree of a matrix $R$ in
order to bound the smallest degree of a polynomial solution to
$R\eta= 0$. Therefore, a bound on the degree of a solution of
Problem~\ref{pb:lin_rel_pseudo_lin_map} can be derived from the
McMillan degree of the associated pseudo-Krylov matrix.
Theorem~\ref{thm:McMillandeg_kronecker} is proved in
Section~\ref{sec:proof_transfer_degree} after recalling some useful
facts and definitions for this work.

\subsection{Determinantal denominators and Smith-McMillan normal form}
\label{sec:det_den}

For $R \in k(x)^{n \times p}$ and a positive integer~$\ell$, we denote
by $\varphi_\ell(R)$ the monic least common denominator of all minors
of $R$ of size at most $\ell$.  We also set $\varphi_0(R) = 1$. The
polynomials $\varphi_\ell(R)$ are called the \emph{determinantal
  denominators} of~$R$~\cite{coppel1974}.  Note that $\varphi_1(R)$ is
the monic least common denominator of the entries of $R$. So one can
write $R = \left(1/\varphi_1(R)\right) \cdot N$ with~$N$ a polynomial
matrix.
By definition, for~$\ell \ge 0$, $\varphi_\ell(R)$ divides
$\varphi_{\ell +1}(R)$ and $\varphi_\ell(R) = \varphi_{\ell+1}(R)$
for~$\ell \ge \rk(R)$. Moreover by a direct expansion of the
determinant, $\varphi_\ell(R)$ divides $\varphi_1(R)^\ell$.  We recall
how determinantal denominators behave with respect to sums, products
and inverse of matrices.

\begin{proposition} \emph{\cite[Thm.~1-2]{coppel1974}}
  \label{prop:det_den_sum_prod}
  If a rational matrix $R$ is the sum $R_1 + R_2$ of two rational
  matrices~$R_1, R_2$, then for all $\ell$, $\varphi_\ell(R)$ divides
  $\varphi_\ell(R_1)\varphi_\ell(R_2)$.  Moreover, if $\varphi_1(R_1)$
  and $\varphi_1(R_2)$ are coprime then equality holds.  Similarly if
  $R$ is the product $R_1R_2$ of two rational matrices $R_1, R_2$, then
  for all~$\ell$,~$\varphi_\ell(R)$ divides
  $\varphi_\ell(R_1)\varphi_\ell(R_2)$.
\end{proposition}

\begin{proposition} \emph{\cite[Thm.~4]{coppel1974}}
  \label{prop:det_den_inverse}
  Let $R$ be a non-singular $m \times m$ rational matrix.  If
  $\det R = c \cdot \alpha / \beta$ with $c \in k \setminus \{0\}$,
  $\alpha, \beta \in k[x]$ monic, then
  $\beta \varphi_m(R^{-1}) = \alpha \varphi_m(R)$.
\end{proposition}

If $R \in k(x)^{n \times p}$, then the matrix $N = \varphi_1(R) \cdot R$ is a
polynomial matrix, hence it has a Smith normal form~\cite{newman1972}
\begin{align*}
  \Gamma =
  \begin{bmatrix}
    \begin{matrix}
      \gamma_1 \\
      & \ddots \\
      && \gamma_r
    \end{matrix} & 0 \\ 0 & 0
  \end{bmatrix} \in k[x]^{n \times p},
\end{align*}
where $r$ is the rank of $N$. The polynomials
$\gamma_1,\dots,\gamma_r$ are the invariant factors of $N$ and they
satisfy~$\gamma_i$ divides $\gamma_{i+1}$ for all $1 \le i < r$.  The
\emph{Smith-McMillan form of $R$} is the
matrix~$S = 1/\varphi_1(R)\cdot \Gamma$~\cite[Sec.~6.5.2, p.~443]
{kailath1980}.  By reducing each rational function on its diagonal, it
is of the form
\begin{align}
  \label{eq:Smith-McMillan_form}
  S =
  \begin{bmatrix}
    \begin{matrix}
      \varepsilon_1 / \psi_1 \\
      & \ddots \\
      && \varepsilon_r / \psi_r
    \end{matrix} & 0 \\ 0 & 0
  \end{bmatrix} \in k(x)^{n \times p},
\end{align}
with $\varepsilon_i$ divides $\varepsilon_{i+1}$ for all
$1 \le i < r$, and $\psi_{i+1}$ divides $\psi_i$. Such a matrix $S$ is
thus uniquely determined.  The denominators $\psi_i$ in $S$ are
strongly connected to the determinantal denominators of $R$.
Indeed~\cite[Thm.~6]{coppel1974}, for all $1 \le i \le r$,
\begin{align}
  \label{eq:detden_prod_den_smithmcmillan}
  \varphi_i(R) = \psi_1 \cdots \psi_i.
\end{align}

\subsection{Matrix fraction description}
\label{sec:MFD}

We consider representations of a rational matrix
$R \in k(x)^{n \times p}$ of the form
\begin{align}
  \label{eq:matrix_realisation}
  R =  XM^{-1}Y,
\end{align}
where $X,M,Y$ are polynomial matrices of sizes
$n \times m, m\times m, m\times p$ respectively and $M$ is
non-singular.  Such a representation for $R$ is called a
\emph{realisation}~\cite{coppel1974}.
\begin{proposition}
  \emph{\cite[Thm.~10]{coppel1974}}
  \label{prop:det_den_realisation}
  Let $R = XM^{-1}Y$ be a realisation of $R$. For all $\ell \ge 0$,
  $\varphi_\ell(R)$ divides~$\det M$.
\end{proposition}
As special cases of realisations, we say that $R = N D^{-1}$ with
$N \in k[x]^{n \times p }, D \in k[x]^{p \times p}$ non-singular is a
\emph{right matrix fraction description}~(MFD), and similarly,
$R = D_L^{-1}N_L$ with
$N_L \in k[x]^{n \times p }, D_L \in k[x]^{n \times n}$ non-singular
is a \emph{left} MFD. In both cases, we say that $D$ or $D_L$ is the
denominator of the description.  Realisations and MFDs are not unique.
For example, for any non-singular $p \times p$ polynomial matrix $V$,
one has $R = NV (DV)^{-1}$.

Recall that a matrix $U \in k[x]^{n\times n}$ is unimodular if its
determinant is a nonzero element in $k$.  Let~$A,B$ be $n \times m$
and $h \times m$ polynomial matrices. A matrix
$G \in k[x]^{m\times m}$ is a \emph{common right divisor} of $A$ and
$B$ if there exist polynomial matrices $A_1, B_1$ such that $A = A_1G$
and $B = B_1G$. The matrices $A$ and $B$ are said to be \emph{right
  relatively prime} if their only common right divisors are
unimodular. Similarly, one can define \emph{common left divisors} and
\emph{relatively left prime} matrices by taking transposes in the
above definitions.  Now a right (resp. left) MFD $R = ND^{-1}$ is said
to be \emph{irreducible} if $N,D$ are right (resp. left) relatively
prime.
\begin{proposition}
  \emph{\cite[Thm.~10]{coppel1974}}
  \label{prop:det_den_MFD} A matrix fraction description of
  $R \in k(x)^{n \times p}$ with denominator $D$ (either right or
  left) is irreducible if and only if
  $\varphi_\ell(R) = \varphi_\ell(D^{-1})$ for all $\ell \ge 0$.
\end{proposition}

\subsection{McMillan degree}
\label{sec:mcmillan-degree}

A matrix $R(x) \in k(x)^{n \times p}$ is called \emph{proper} if
$R(x)$ admits a finite limit when $x$ tends to infinity.  In other
words, for every entry of $R$, its numerator has degree less or equal
than its denominator.  A strictly proper matrix tends to $0$ at infinity.

When $R$ is proper, the \emph{McMillan degree of $R$}, denoted
$\degMM(R)$, is defined as the sum of the degrees of the denominator
polynomials in its Smith-McMillan form~\cite[p.~444]{kailath1980}.
By~(\ref{eq:detden_prod_den_smithmcmillan}),
\begin{align}
  \label{eq:mcmillandef_def_proper}
  \degMM(R) = \deg(\varphi_r(R)),
\end{align}
where $r$ is the rank of $R$. For arbitrary matrix
$R \in k(x)^{n \times p}$, one can always decompose $R = R_o + W$
where $R_o$ is strictly proper and $W$ is a polynomial matrix. Then,
$W(1/x)$ is proper and one can define~\cite[p.~466]{kailath1980}
\begin{align}
  \label{eq:mcmillandeg_def}
  \degMM(R) = \degMM(R_o) + \degMM(W(1/x)).
\end{align}

Also, $\degMM(R) = \sum_{\alpha} g_\alpha$~\cite[Cor.~2]{DuHa1963}
where $\alpha$ ranges over all the poles of $R$, infinity included,
and~$g_\alpha$ is the maximum order to which the pole $\alpha$ occurs
in a minor of $R$ of arbitrary size.  In particular, for polynomial
matrices $M$ of rank $r$, $\degMM(M)$ is the maximum of the degrees of
the~$i \times i$ minors of $M$ for all $i \le r$.  We also define the
\emph{minor degree} of a polynomial matrix $M$ to be the maximum of
the degrees of minors of $M$ of size \emph{exactly} $r$ and denote it
$\Minordeg(M)$.  By a slight abuse of notation, the \emph{McMillan
  degree} sometimes refers the \emph{minor degree} in the literature
as in~\cite{StoVi2005report}.  Yet in this work, we distinguish these
two notions of degrees.

By a result of Kalman, one can always reduce to proper rational
matrices thanks to a suitable change of variables.

\begin{proposition}
  \label{prop:mcmillandeg_proper_mobius}
  \emph{\cite[Prop.~11]{kalman1965}} For any rational matrix
  $R \in k(x)^{n \times p}$, there exists a
  transformation~$\mu(x) = (\alpha x + \beta)/(\gamma x + \delta)$
  where $\alpha,\beta,\gamma,\delta \in k$,
  $c = \alpha \delta - \beta \gamma \neq 0$, $\gamma \neq 0$,
  such that $R(\mu(x))$ is proper and
  $\degMM(R(x)) = \degMM(R(\mu(x)))$. The result still holds for a
  finite family of matrices~$(R_1,\dots, R_\ell)$, there exists $\mu$
  such that for all $1\le i \le \ell$, $R_i(\mu)$ is proper and
  $\degMM(R_i) = \degMM(R_i(\mu))$.
\end{proposition}
In Proposition~\ref{prop:mcmillandeg_proper_mobius}, the
transformation $\mu$ is an invertible change of variables that sends
all the poles of the matrix, infinity included to finite distinct new
poles.  In particular, $\mu$ moves the point at infinity to the finite
point $-\delta/\gamma$.  For a rational matrix (or vector) $R$, let
$R_\mu$ denote the transformed matrix~$R(\mu(x))$. By~(\ref{eq:mcmillandef_def_proper}), we have
$\degMM(R) = \deg (\varphi_r(R_\mu))$ with $r$ the rank of $R$.  We
further rely on the following properties of McMillan degree.
\begin{proposition}
  \emph{\cite{mcmillan1952II,DuHa1963}}
  \label{prop:mcmillandeg_prod_inv}
  Let $R$ be a rational matrix. If $R$ is a product $R_1 R_2$ or a sum
  $R_1 + R_2$, then $\degMM(R) \le \degMM(R_1) + \degMM(R_2)$. If $R$
  is square and non-singular, then $\degMM(R^{-1}) = \degMM(R)$.
\end{proposition}

When the matrices involved in
Proposition~\ref{prop:mcmillandeg_prod_inv} are proper,
the results directly follow from
Propositions~\ref{prop:det_den_sum_prod}
and~\ref{prop:det_den_inverse} on determinantal denominators. In the
general case, one reduces to the proper case by a change of variable
of the shape of Proposition~\ref{prop:mcmillandeg_proper_mobius}.
Therefore, up to a change of variables, the McMillan degree of a matrix
$R$ is given by the
degree of the 
largest determinantal denominator of the matrix, which, by Propositions~\ref{prop:det_den_realisation}
and~\ref{prop:det_den_MFD}, can be estimated
via a realisation or a description of $R$.

\subsection{Proof of Theorem~\ref{thm:McMillandeg_kronecker}}
\label{sec:proof_transfer_degree}

First note that the second inequality in~(\ref{eq:kronecker_mcmillan})
is a direct consequence of the first one since
$v_1\le \cdots \le v_{p-r}$. To prove the first inequality, we first
reduce to the case where $R$ is proper using
Proposition~\ref{prop:mcmillandeg_proper_mobius}.  Next, when $R$ is
proper, the result is obtained from~\cite[Thm.~3.3]{StoVi2005issac}
that we apply to the numerator matrix $A \in k[x]^{n \times p}$ of a
left MFD $B^{-1}A$ of $R$, since $R$ and $A$ share the same nullspace.

Let $N$ be a minimal basis of the nullspace of $R$ with column degree
$v_1 \le \cdots \le v_{p-r}$.  Let~$\mu(x)$ be a transformation as in
Proposition~\ref{prop:mcmillandeg_proper_mobius} and let
$u(x) = \gamma x + \delta$ be the denominator of $\mu$.  Then, we
have~$R_\mu N_\mu = 0$ and $N_\mu = N^* D^{-1}$ with $N^*$ a polynomial
matrix with column degree~$v_1^*\le \cdots\le v_{p-r}^*$ satisfying
$v_i^* \le v_i$ for all $i$ and $D$ the diagonal matrix
$(u^{v_1},\dots,u^{v_{p-r}})$. Since $R_\mu N^* = 0$, $N^*$ is a basis
of the nullspace of $R_\mu$.  By the same reasoning, since $\mu$ is
invertible, from any minimal basis of $R_\mu$ with column degree
$\nu_1\le \cdots \le \nu_{p-r}$, one can construct a basis of~$R$ with
column degree~$\nu_1^*\le\cdots\le\nu_{p-r}^*$ satisfying
$\nu_i^* \le \nu_i$ for all $i$. Therefore $N^*$ is a minimal basis of
$R_\mu$, and thus $R$ and $R_\mu$ share the same Kronecker indices.  So, by
Proposition~\ref{prop:mcmillandeg_proper_mobius}, one can assume that
$R$ is proper.

Let $B^{-1}A$ be an irreducible left MFD of $R$ and suppose that $B$
is row-reduced, \ie $B$ is such that
$\deg (\det B) = \sum_{i=1}^n b_i$ where~$b_i$ is the degree of the
row $i$ of $B$.  It is always possible
to find such a description~(see e.g.,~\cite[Sec.~6.7.2,
p.~481]{kailath1980}).  Since $R\eta = 0$ if and only if $A \eta = 0$,
$A$ and $R$ have the same Kronecker indices.
By~\cite[Thm.~3.3]{StoVi2005issac}, we have
\begin{align*}
  \sum_{i=1}^{p-r}v_i \le \Minordeg(A).
\end{align*}
Now let $a_1,\dots,a_n$ denote the degrees of the rows of
$A$. Clearly, we have $ \Minordeg(A) \le \sum_{i=1}^n a_i $, and
because $B^{-1}A$ is proper, we have $a_i \le b_i$ for
$ 1 \le i \le n$. Next, since $B$ is row-reduced, we get
$\deg (\det B) = \sum_{i=1}^n b_i \ge \sum_{i=1}^{p-r} v_i$.  The
properness of $R$ yields
$\degMM(R) = \deg \varphi_r(R) = \deg \varphi_n(R) = \deg
\varphi_n(B^{-1})$ by Proposition~\ref{prop:det_den_MFD}, since the
description is irreducible. Finally, by
Proposition~\ref{prop:det_den_inverse}, $\varphi_n(B^{-1}) = c\det B$
with $c \in k \setminus \{0\}$, therefore $\degMM(R) = \deg(\det B)$,
whence the result. \qed

\section{Degree bounds for pseudo-Krylov systems}
\label{sec:deg_bounds_pseudo_krylov}

In this section, our main result is established: we bound the degrees
of the polynomial coefficients in the solutions of
Problem~\ref{pb:lin_rel_pseudo_lin_map} for pseudo-linear maps of the
form $\theta = p(x)\DP_x +T$ with $p$ a polynomial of degree at most
$1$ or $\theta = T\cdot \sigma_x$.  We also derive an order-degree
curve by relaxing the minimality on the order in
Problem~\ref{pb:lin_rel_pseudo_lin_map} and allowing over-determined
pseudo-Krylov systems.

The degree bound is obtained from
Theorem~\ref{thm:McMillandeg_kronecker} by bounding the McMillan
degree of the matrix of the pseudo-Krylov
system~(\ref{eq:lin_rel_pseudo_lin_map}).

\begin{theorem}
  \label{thm:bd_mcmillan_pseudoKrylov}
  Let $T \in k(x)^{n \times n}$, $\theta = p(x)\DP_x + T$ with
  $p \in k[x]$ of degree at most $1$ or $\theta = T \cdot \sigma_x$,
  and~$a \in k[x]^n$ of degree $d_a$.  Let
  $K \in k(x)^{n \times (m+1)}$ be the pseudo-Krylov matrix $ K =
  \begin{bmatrix}
    a & \theta a & \cdots & \theta^{m}a
  \end{bmatrix},
  $
  and let $\rho$ denote its rank.
  Then,
  \begin{align*}
    \degMM(K) \le \rho d_a + m \degMM(T).
  \end{align*}
\end{theorem}

\noindent The proof of Theorem~\ref{thm:bd_mcmillan_pseudoKrylov} is
developed in the rest of the section.  As a consequence, we obtain
Theorem~\ref{thm:degree_bound_pseudo_krylov}.

\begin{proof}[Proof of Theorem~\ref{thm:degree_bound_pseudo_krylov}]
  Let $K=
  \begin{bmatrix}
    a & \theta a & \cdots & \theta^m a
  \end{bmatrix} \in k(x)^{n \times (m+1)}$. Note that $\rho$ is the rank of $K$.
  By Theorem~\ref{thm:McMillandeg_kronecker},
  the right Kronecker index $v_1$ of $K$
  satisfies
  $ v_1 \le \degMM(K)/{(m+1-\rho)}$,
  and by Theorem~\ref{thm:bd_mcmillan_pseudoKrylov},
  there exists an element in the nullspace of $K$ of degree at most
  $1/( m+1-\rho) \cdot (\rho d_a + m \degMM(T))$.
  The last claim follows by setting $m =\rho$.
\end{proof}

The general sketch for proving
Theorem~\ref{thm:bd_mcmillan_pseudoKrylov} is the following: if both
$T$ and $K$ are proper matrices, then
by~(\ref{eq:mcmillandef_def_proper}), their McMillan degree is
completely determined by their determinantal denominators.  In
Section~\ref{sec:det_den_pseudo_krylov}, we study the determinantal
denominators of pseudo-Krylov matrices.~Finally in
Section~\ref{sec:bounds_McMillan_degree}, we prove
Theorem~\ref{thm:bd_mcmillan_pseudoKrylov} by reducing the general
case to the proper case using a change of variable and
Proposition~\ref{prop:mcmillandeg_proper_mobius}.  However after
applying the change of variable $\mu$ in $K$, the obtained matrix
$K_\mu$ is no longer a pseudo-Krylov matrix satisfying the assumptions
of Theorem~\ref{thm:bd_mcmillan_pseudoKrylov}.  To circumvent this
difficulty, in Section~\ref{sec:det_den_pseudo_krylov}, we study the
determinantal denominators of a broader class of matrices than the
pseudo-Krylov matrices in Theorem~\ref{thm:bd_mcmillan_pseudoKrylov}
(see Propositions~\ref{prop:det_den_krylov_diffeq}
and~\ref{prop:det_den_krylov_rec}).  In particular, in the
differential case, this leads us to study the determinantal
denominators of pseudo-Krylov matrices built from an operator
$\theta = p(x) \DP_x + T$ with $p(x)$ a polynomial of \emph{arbitrary
  degree}.

\subsection{Determinantal denominators of pseudo-Krylov matrices}
\label{sec:det_den_pseudo_krylov}

For $T \in k(x)^{n \times n}$, we study the determinantal denominators
of pseudo-Krylov matrices of the following shape:
\begin{align}
  \label{eq:pseudo_Krylov_matrix}
  K =
  \begin{bmatrix}
    \theta^{s_1} u_1 & \cdots & \theta^{s_m}u_m  
  \end{bmatrix}
  \in k(x)^{n \times m},
\end{align}
with $u_1, \dots, u_m \in k[x]^n$, $s_1,\dots,s_m$ being non-negative
integers, and $\theta$ being the pseudo-linear map~$T\sigma_x$
or~$p(x)\DP_x +T$, where in this section, $p(x)$ is a
polynomial of \emph{arbitrary degree}.

\subsubsection{Differential case}
\label{sec:det_den_diff_case}

We start by the differential case with $\theta = \DP + T$, where $\DP$
is any derivation on $k(x)$ of the form~$p(x)\DP_x$ where
$p(x) \in k[x]$.  An elementary reasoning by induction shows the
following.
\begin{lemma}
  \label{lem:ind_pow_theta_diff_case}
  Let $T = D^{-1}N \in k(x)^{n\times n}$ be a description,
  $\theta = \DP +T$ and $\theta_1 = \DP + (N-\DP(D))D^{-1}$.  For all
  $s \ge 0$, $\theta^s = \DP^s+ D^{-1} \Omega_s$ with
  $ \Omega_s = \sum_{i=0}^{s-1} \theta_1^iN\DP^{ s-1-i}$.
\end{lemma}

\begin{proof}
  The results holds for $s=0$. Suppose it holds for $s-1 \ge 0$, then
  by induction
  \begin{align*}
    \theta^s  &= \theta \theta^{s-1} = \left(\DP + D^{-1}N \right)
                \left( \DP^{s-1} + D^{-1} \Omega_{s-1} \right),
  \end{align*}
  Then, since $\DP(D^{-1}) = - D^{-1} \DP(D) D^{-1}$,
  \begin{align*}
    \theta^s  &= \DP^{s} + D^{-1}N\DP^{s-1} + D^{-1}ND^{-1}\Omega_{s-1}
                - D^{-1}\DP(D)D^{-1}\Omega_{s-1} +
                D^{-1}\DP \Omega_{s-1} \\
              &= \DP^{s} + D^{-1}\left(N\DP^{s-1} + \left( (N-\DP(D))D^{-1}+ \DP\right)
                \Omega_{s-1} \right) =
                \DP^{s} + D^{-1}\left(N\DP^{s-1} + \theta_1\Omega_{s-1} \right) \\
              &=    \DP^{s}+ D^{-1}\Omega_s. \qedhere
  \end{align*}
\end{proof}

\begin{proposition}
  \label{prop:det_den_krylov_diffeq}
  Let $s_1 \le \cdots \le s_m$ be non-negative integers,
  $T \in k(x)^{n\times n}$ and $\theta = \DP +T$. For any vectors of
  polynomials $u_1, \dots, u_m \in k[x]^n$, the pseudo-Krylov matrix
  \begin{align*}
    K =
    \begin{bmatrix}
      \theta^{s_1} u_1 & \cdots & \theta^{s_m}u_m  
    \end{bmatrix}
    \in k(x)^{n \times m},
  \end{align*}
  satisfies $\varphi_\ell(K)$ divides $\varphi_\ell(T)^{s_m}$ for all
  $\ell \ge 0$.
\end{proposition}

\begin{proof}
  We prove that any minor of size $\ell$ of $K$ can be written with
  denominator $\varphi_\ell(T)^{s_m}$. It suffices to prove the result
  for $\ell = m$ since one can delete in $K$ the $m-\ell$ columns that
  are not selected in the minor.
  We proceed by induction on $m+ s_m \ge 1$ with $m \ge 1$.  For
  $m= 1$, the result holds since~$\theta^{s_m}u_m$ can be written with
  denominator $\varphi_1(T)^{s_m}$.  For $s_m =0$,
  we have $K \in k[x]^{n \times m}$, and thus $\varphi_m(K) = 1$.  Now suppose
  that $m \ge 2$ and $s_m \ge 1$.  First, if $s_1 = 0$, the first
  column of $K$ has polynomial entries.  Let $\k$ be a minor of $K$ of
  size $m$, one can expand $\k$ along the first column of~$K$, thus it
  is a polynomial linear combination of minors of size $m-1$ of the $m-1$ last
  columns of~$K$ with rational coefficients.  By induction, it can
  be written with denominator $\varphi_{m-1}(T)^{s_m}$ which
  divides~$\varphi_{m}(T)^{s_m}$. Finally if~$s_1 \ge 1$, let $T= D^{-1}N$ be
  an irreducible left MFD of $T$. Then, by
  Lemma~\ref{lem:ind_pow_theta_diff_case} for $1 \le j \le m$,
  \begin{align*}
    \theta^{s_j}u_j = u_j^{(s_j)} + D^{-1}
    \Omega_{s_j}(u_j),
  \end{align*}
  where $\Omega_s = \sum_{i=0}^{s-1} \theta_1^iN\DP^{ s-1-i}$ for all
  $s \ge 0$, $\theta_1$ is the pseudo-linear map $\DP + T_1$ with
  $T_1 = (N-\DP(D))D^{-1}$. So, we get
  \begin{align*}
    K =
    \begin{bmatrix}
      u_1^{(s_1)} & \cdots & u_m^{(s_m)}
    \end{bmatrix} + D^{-1} \cdot K_1,
  \end{align*}
  where the $j$-th column of $K_1 \in k(x)^{n \times m}$ is 
  $\Omega_{s_j}(u_j)$. By Propositions~\ref{prop:det_den_sum_prod}
  and~\ref{prop:det_den_MFD}, $\varphi_m(K)$ divides $\varphi_m(D^{-1})
  \varphi_m(K_1) = \varphi_m(T) \varphi_m(K_1)$.
  Now, by multi-linearity of the determinant, any minor of size~$m$ of
  $K_1$ is the sum of the same minors of the matrices
  \begin{align*}
    \bar K =
    \begin{bmatrix}
      \theta_1^{i_1}(Nu_1^{(s_1-1-i_1)}) & \cdots &  \theta_1^{i_m}(Nu_r^{(s_m-1-i_m)})
    \end{bmatrix},
  \end{align*}
  for $0 \le i_j < s_j$ and $1 \le j \le m$. By induction,
  $\varphi_m(\bar K)$ divides $\varphi_m(T_1)^{\max_j(i_j)}$ and thus
  all these minors can be written with denominator
  $\varphi_m(T_1)^{s_m-1}$. Since $\DP(D)$ is a polynomial matrix,
  $(N -\DP(D))D^{-1}$ is a description of $T_1$.  Thus by
  Propositions~\ref{prop:det_den_realisation}
  and~\ref{prop:det_den_MFD}, $\varphi_m(T_1)^{s_m-1}$ divides
  $\varphi_m(D^{-1})^{s_m-1} = \varphi_m(T)^{s_m-1}$. Hence,
  $\varphi_m(K)$ divides $\varphi_m(T)^{s_m}$.
\end{proof}

\subsubsection{Shift case}
\label{sec:det_den_shift_case}

We now turn to the shift case with $\theta = T \cdot \sigma_x$.  By
induction, we have the identity
$\theta^s = T \sigma_x(T)\cdots \sigma_x^{s-1}(T)\sigma_x^s$ for all
$s \ge 0$.
This motivates the following result.

\begin{proposition}
  \label{prop:det_den_krylov_rec}
  Let $0 \le s_1 \le \cdots \le s_m$ be non-negative integers,
  $T_0,\dots, T_{s_m - 1} \in k(x)^{n \times n}$ be~$s_m$ matrices of
  rational functions and for $0 \le i\le s_m$, let
  $T^{(i)} = T_0 \cdots T_{i-1}$.  For any vectors of polynomials
  $u_1, \dots, u_m \in k[x]^n$, the matrix
  \begin{align*}
    K =
    \begin{bmatrix}
      T^{(s_1)} u_1 & \cdots & T^{(s_m)}u_m  
    \end{bmatrix}
    \in k(x)^{n \times m},
  \end{align*}
  satisfies $\varphi_\ell(K)$ divides
  $\varphi_\ell(T_0)\cdots \varphi_\ell(T_{s_m-1})$ for all
  $\ell \ge 0$.
\end{proposition}
\begin{proof}
  As in the proof of Proposition~\ref{prop:det_den_krylov_diffeq} it
  suffices to prove the result for $\ell = m$, that is, any minor of
  size $m$ of $K$ can be written with denominator
  $\varphi_m(T_0)\cdots \varphi_m(T_{s_m-1})$. We again proceed by
  induction on $m + s_m \ge 1$ with $m \ge 1$.  If $m=1$,
  $K = T^{(s_m)}u_m = T_0 \cdots T_{s_m -1 }u_m$, and it can be
  written with denominator
  $\varphi_1(T_0)\cdots \varphi_1(T_{s_m-1})$.  If $s_m = 0$, $K$ is a
  polynomial matrix whose columns are~$u_1,\dots, u_m$, thus
  $\varphi_m(K) = 1$ and the result holds.  Now we can suppose that
  $m \ge 2$ and~$s_m \ge 1$.  If $s_1 = 0$, we can conclude by
  induction by expanding any minor of size $m$ along its first column.
  Any minor of size $m$ of $K$ is a polynomial linear combination of
  minors of size $m-1$ of the~$m-1$ last columns of $K$.  So, it can
  be written with denominator
  $\varphi_{m-1}(T_0)\cdots \varphi_{m-1}(T_{s_m-1})$, and the result
  holds since $\varphi_{m-1}(R)$ divides $\varphi_m(R)$ for any matrix
  $R$.  Finally, suppose that $s_1 \ge 1$. We have~$T^{(s)}
  = T_0 \cdot T^{(1,s)}$ for all~$1 \le s \le s_m$, with
  $ T^{(1,s)} = T_1 \cdots T_{s-1}$.  Therefore, we obtain~$K = T_0 \cdot K_1$
  with
  \begin{align*}
    K_1 =
    \begin{bmatrix}
      T^{(1,s_1)} u_1 & \cdots & T^{(1,s_m)}u_m)
    \end{bmatrix}.
  \end{align*}
  Thus, $\varphi_m(K)$ divides $\varphi_m(T_0)\varphi_m(K_1)$ by
  Proposition~\ref{prop:det_den_sum_prod}. We conclude by induction,
  since $\varphi_m(K_1)$ divides
  $\varphi_m(T_1)\cdots \varphi_m(T_{s_m-1}) $.
\end{proof}

\subsection{Bounds on McMillan degree}
\label{sec:bounds_McMillan_degree}

In this section, we provide bounds on the McMillan degree of
pseudo-Krylov matrices and prove
Theorem~\ref{thm:bd_mcmillan_pseudoKrylov}.

\subsubsection{Differential case}
\label{sec:McMillan_pseudoKrylov_diff}

\begin{lemma}
  \label{lem:pseudo_krylov_diffeq_mobius}
  Let $T \in k(x)^{n \times n}$ and consider the pseudo-linear map
  $\theta = p(x)\DP_x + T$ with $p$ a polynomial of degree at most 1.
  Let $a\in k(x)^n$, and $j\ge0$.  Then, for all transformations
  $\mu(x)$ of the shape of
  Proposition~\ref{prop:mcmillandeg_proper_mobius},
  \begin{align*}
    (\theta^j a)(\mu(x)) = \bar \theta^j(a_\mu),
  \end{align*}
  with $a_\mu = a(\mu(x))$ and $\bar \theta = \Delta + T_\mu$
  where $\Delta$ is the derivation $ (p_\mu/\mu') \cdot \DP_x$ over
  $k(x)$ with $T_\mu = T(\mu(x))$ and $p_\mu = p(\mu(x))$.  Moreover, letting $u$ the
  denominator of $\mu$, if $a \in k[x]^{n}$ and has degree $d_a$, then
  there exists $a^* \in k[x]^n$ of degree at most $d_a$ such that for
  all $j \ge 0$,
  \begin{align}
    \label{eq:triangular_thetabaramu}
    \bar \theta^j a_\mu  = 1/u^{d_a} \sum_{i=0}^j {p_{i,j}} \bar \theta^i a^*,
  \end{align}
  where 
  $p_{i,j} = \binom{j}{i} u^{d_a} \Delta^{j-i}(1/u^{d_a})$ is a \emph{polynomial}.
\end{lemma}

\begin{proof}
  The first claim holds for $j=0$. Suppose it holds for $j \ge 0$.
  Because, $\theta^{j+1}a= T \cdot \theta^j a + p (\theta^j a)'$,
  then
  $\left(\theta^{j+1}a\right)(\mu) = T_\mu \left( (\theta^j a)(\mu)
  \right) + p_\mu (\theta^j a)'(\mu)$.  However, by the chain rule
  $ (\theta^j a)'(\mu)= 1/\mu' \left((\theta^j a)(\mu)\right)'$.
  Thus,
  $\left(\theta^{j+1}a\right)(\mu) = \bar
  \theta \left((\theta^j a)(\mu)\right) = \bar \theta^{j+1}(a_\mu)$,
  by induction.
  
  Next, let $\mu = (\alpha x  + \beta)/(\gamma x + \delta)$
  with $\alpha, \beta, \gamma,\delta \in k$, $c= \alpha \delta - \beta \gamma \neq 0$,
  $\gamma \neq 0$ and $u = \gamma x + \delta$. Then, $\mu' = c / u^2$.
  Since $p$ is of degree at most 1, we have $p_\mu / \mu' = u^2 p_\mu/c$ is a polynomial
  and is divisible by $u$.  Therefore, by induction on $\ell$,
  $\Delta^\ell(1/u^{d_a})$ can be written with denominator $u^{d_a}$
  and thus, $p_{i,j}$ is a polynomial for all~$i,j$.  Besides, we
  have $a_\mu = a^*/u^{d_a}$ with $a^{*}$ a vector of polynomials of
  degree at most $d_a$. Then~(\ref{eq:triangular_thetabaramu}) holds
  for $j=0$. Finally, suppose that it holds for $j \ge 0$. Then by induction, we have 
  \begin{align*}
    \bar \theta^{j+1}(a_\mu)
    &= \sum_{i=0}^j \left(\binom{j}{i}\Delta^{j-i}(1/u^{d_a})
      \bar \theta^{i+1} a^*
      + \binom{j}{i}\Delta^{j-i+1}(1/u^{d_a}) \bar \theta^{i} a^*\right) \\
    &= \sum_{i=0}^{j+1} \left[\binom{j}{i-1} + 
      \binom{j}{i}  \right] \Delta^{j-i+1}(1/u^{d_a})\bar \theta^{i} a^* 
      = \sum_{i=0}^{j+1} \binom{j+1}{i} \Delta^{j+1-i}(1/u^{d_a})
      \bar \theta^{i} a^*,
  \end{align*}
  whence the result is derived.
\end{proof}

\begin{proof}[Proof of
  Theorem~\ref{thm:bd_mcmillan_pseudoKrylov} for
  $\theta = p\DP_x + T$]
  Suppose first that both $T$ and $K$ are proper rational matrices.
  In particular, $a$ is a constant vector and $d_a = 0$. Also
  by~(\ref{eq:mcmillandef_def_proper}), we have
  $\degMM(T) = \deg \varphi_n(T)$
  and~$\degMM(K) = \deg \varphi_\rho(K) = \deg \varphi_n(K)$.  However,
  by Proposition~\ref{prop:det_den_krylov_diffeq}, $\varphi_n(K)$
  divides $\varphi_n(T)^m$. Hence $\degMM(K) \le m \degMM(T)$.  So
  the result holds in the case where both $T$ and $K$ are proper.
  
  In the general case, we can reduce it to the proper case. By
  Proposition~\ref{prop:mcmillandeg_proper_mobius}, there exists a
  transformation~$\mu(x)$ with denominator $u(x)$ of degree 1,
  such that both $T_\mu = T(\mu(x))$ and $K_\mu = K(\mu(x))$ are
  proper rational matrices and
  $\degMM(T) = \degMM(T_\mu) = \deg \varphi_n(T_\mu)$ and
  $\degMM(K) = \degMM(K_\mu) = \deg \varphi_\rho(K_\mu)$.  By
  Lemma~\ref{lem:pseudo_krylov_diffeq_mobius}, $K_\mu$ is the
  pseudo-Krylov matrix $
  \begin{bmatrix}
    a_\mu & \bar \theta a_\mu & \cdots & \bar \theta ^m a_\mu
  \end{bmatrix}$, where $a_\mu = a(\mu(x)) \in k(x)^n$ and
  $\bar \theta$ is the pseudo-linear map $\Delta + T_\mu$
  with $\Delta$ the derivation $p_\mu /\mu' \cdot \DP_x$ over $k(x)$ with $p_\mu = p(\mu)$.
  Note also that $p_\mu/\mu'$ is a polynomial
  (see the proof of Lemma~\ref{lem:pseudo_krylov_diffeq_mobius}). 
  Moreover by~(\ref{eq:triangular_thetabaramu}),
  there exists $a^* \in k[x]^n$ which is a vector of \emph{polynomials}
  of degree at most $d_a$ such that 
  \begin{align*}
    K_\mu = K^*\cdot  \frac{1}{u^{d_a}} P,
  \end{align*}
  where $P$ is an upper-triangular \emph{polynomial} matrix and
  $K^* =
  \begin{bmatrix}
    a^* &  \bar \theta a^* & \cdots & \bar \theta ^m a^*
  \end{bmatrix}$.
  Since $P$ is a polynomial matrix, $\varphi_\rho(1/u^{d_a}P)$ divides
  $u^{\rho d_a}$, thus by Proposition~\ref{prop:det_den_sum_prod},
  $\varphi_\rho(K_\mu)$ divides $u^{\rho
    d_a}\varphi_\rho(K^*)$. However, $K^*$ is a pseudo-Krylov matrix
  associated to the pseudo-linear map $\bar \theta$ and initial
  \emph{polynomial} vector $a^*$, so by
  Proposition~\ref{prop:det_den_krylov_diffeq},
  $\varphi_\rho(K^*) = \varphi_n(K^*)$ divides
  $\varphi_n(T_\mu)^m$. In particular, it has degree at most
  $m \deg \varphi_n(T_\mu) = m \degMM(T)$. Hence since $\deg u = 1$,
  we conclude that
  $\degMM(K) = \deg \varphi_\rho(K_\mu) \le \rho d_a + m \degMM(T)$.
\end{proof}

\subsubsection{Shift case}
\label{sec:McMIllan_pseudoKrylov_shift}

\begin{lemma}
  \label{lem:degMM-shift}
  Let $T \in k(x)^{n \times h}$ be an arbitrary matrix of rational
  functions.  Then, $\degMM(\sigma_x(T)) = \degMM(T)$.
\end{lemma}

\begin{proof}
  Write $T= W +T_o$ where $W$ is a polynomial matrix and $T_o$ is
  proper.  Then, $\sigma_x(T) = \sigma_x(W) + \sigma_x(T_o)$ with
  $\sigma_x(T_o)$ proper.
  Now for any minor of $T_o$ of the form $A/B$ with $A,B \in k[x]$ coprime,
  the same minor of $\sigma_x(T_o)$ is $\sigma_x(A)/\sigma_x(B)$, and
  $\sigma_x(A)$ and $\sigma_x(B)$ are coprime polynomials. Therefore,
  for all $\ell \ge 0$, $\varphi_\ell(\sigma_x(T_o)) = \sigma_x(\varphi_\ell(T_o))$.
  Thus, by~(\ref{eq:mcmillandef_def_proper}),
  $\degMM(T_o) = \degMM(\sigma_x(T_o))$.  Also,
  $\degMM(W) = \degMM(W(1/x))$ is the maximum of the degrees of any
  minor of $W$.  Since $\deg(\sigma_x(p)) = \deg(p)$ for any polynomial
  $p$, then $\degMM(W) = \degMM(\sigma_x(W))$, and
  by~(\ref{eq:mcmillandeg_def}), $\degMM(T) = \degMM(\sigma_x(T))$.
\end{proof}

\begin{proof}[Proof of
  Theorem~\ref{thm:bd_mcmillan_pseudoKrylov} for
  $\theta = T\cdot \sigma_x$]
  Recall that
  $\theta^s = T \sigma_x(T)\cdots \sigma_x^{s-1}(T)\sigma_x^s$ for all
  $s \ge 0$. Again, suppose first that both $T$ and $K$ are proper
  rational matrices, then so are $\sigma_x^{i}(T)$ for all $i$.  In
  particular, $d_a = 0$. By Proposition~\ref{prop:det_den_krylov_rec},
  $\varphi_\rho(K)$ divides
  $\varphi_\rho(T)\cdots \varphi_\rho(\sigma_x^{m-1}(T))$, and thus
  $\degMM(K) = \deg \varphi_\rho(K) \le \sum_{i=0}^{m-1} \deg (
  \varphi_\rho(\sigma_x^{i}(T))) \le \sum_{i=0}^{m-1}
  \degMM(\sigma_x^{i}(T)) = m\degMM(T)$ by
  Lemma~\ref{lem:degMM-shift}.

  We again reduce the general case to the proper case.  By
  Proposition~\ref{prop:mcmillandeg_proper_mobius}, there exists a
  transformation $\mu(x)$ with denominator $u$ of degree 1,
  such that
  $K_\mu = K(\mu(x))$ and $T_{\mu + i} = (\sigma_x^i(T))(\mu(x))$ for
  $0 \le i < m$ are all proper rational matrices and
  $\degMM(K) = \degMM(K_\mu) = \deg \varphi_\rho(K_\mu)$
  and~$\degMM(\sigma_x^i(T)) = \degMM(T_{\mu+i}) = \deg
  \varphi_n(T_{\mu+i})$. For all $j \ge 0$, we have
  \begin{align*}
    (\theta^j a)(\mu(x)) = T_\mu \cdot T_{\mu+1} \cdots T_{\mu + j-1} a(\mu(x) + j)
    = 1/u^{d_a} \cdot T_\mu  \cdots T_{\mu + j-1}\cdot a_j^*,
  \end{align*}
  where $a_j^*$ is a vector of polynomials of degree at most $d_a$.
  Therefore,
  \begin{align*}
    K_\mu = 1/u^{d_a} \cdot K^*,
  \end{align*}
  where $K^* \in k(x)^{n \times (m+1)}$ whose column $j$ for
  $0 \le j \le m$ is $T_\mu \cdots T_{\mu + j-1} \cdot a_j^*$.  So,
  $\varphi_\rho(K_\mu)$ divides $u^{\rho d_a}\varphi_\rho(K^*)$.
  However by Proposition~\ref{prop:det_den_krylov_rec},
  $\varphi_\rho(K^*)$ divides
  $\varphi_{\rho}(T_\mu) \cdots \varphi_\rho( T_{\mu + m-1})$ and thus
  divides $\varphi_{n}(T_\mu) \cdots \varphi_n( T_{\mu + m-1})$ since
  $\rho \le n$.  Finally, using $\deg u =1$, we conclude that
  $\degMM(K) = \deg \varphi_\rho(K_\mu) \le \rho d_a +
  \sum_{i=0}^{m-1} \degMM(T_{\mu+i}) = \rho d_a + \sum_{i=0}^{m-1}
  \degMM(\sigma_x^i(T)) = \rho d_a + m\degMM(T)$ by
  Lemma~\ref{lem:degMM-shift}.
\end{proof}

This completes the proof of
Theorem~\ref{thm:bd_mcmillan_pseudoKrylov}, whence
Theorem~\ref{thm:degree_bound_pseudo_krylov} is also proved.  In the next
sections, Theorem~\ref{thm:degree_bound_pseudo_krylov} is applied to several
problems involving differential and recurrence operators.  For each
problem, a degree bound is derived on the coefficients of a solution
and this bound is compared with the literature.
Sections~\ref{sec:lclm}-\ref{sec:closure_polynomials} deal with
closure properties for Ore operators (LCLM, symmetric product,...).
For the sake of clarity, these operations are presented in the special
case of usual differential operators in $\DP_x$. Yet all the
discussions and results remain valid for differential operators in
$\Delta = p(x)\DP_x$ with $p \in k[x]$ of degree at most 1 and
recurrence operators in~$\sigma_x$.

\section{Least common left multiple}
\label{sec:lclm}

We start with the closure by sums for $D$-finite
functions. Let~$L_1,
\dots, L_s \in k[x]\left<\DP_x\right>$,
we want~$L = \LCLM(L_1,\dots,L_s)$  a least common left
multiple of the $L_i$'s, that is, a minimal order operator that annihilates
all the sums of solutions of the $L_i$'s.

\begin{theorem}
  \label{thm:degree_bound_LCLM}
  Let $L_1,\dots, L_s \in k[x]\left<\DP_x\right>$ be of respective orders
  $r_1, \dots, r_s$, and degrees at most $d$ in~$x$. Then, there exists
  an $LCLM$ $L$ of $L_1, \dots, L_s$  of order $\rho \le \sum_{i=1}^s r_i$ and
  degree at most $s\rho d$.  For all~$m \ge \rho$, there exists a
  nonzero common left multiple of order $m$ and degree at most
  $smd/(m-\rho+1)$.
\end{theorem}

For least common left multiples of $s$ operators of order at most $r$ and
degree at most $d$, our bound behaves like $ds^2r$. We retrieve the leading term
of the best known bound for LCLMs~\cite[Thm.~6]{BoChSaLi2012}.
Besides, for all $m \ge \rho$, there exists common left
multiples of $L_1,\dots,L_s$ of order $m$ and degree at most
$msd/(m-\rho+1)$. By Theorem~\ref{thm:degree_bound_LCLM},
the arithmetic size of a common left multiple is minimised
for~$m =2(\rho-1)$, so there exists a common left multiple of
arithmetic size at most~$4(\rho-1)sd \le 4 s^2 d r$ which is a small
improvement over the bound
$O(s^2(d+r )^2)$~\cite[Thm.~10]{BoChSaLi2012}.

\subsection{Reduction to Problem~\ref{pb:lin_rel_pseudo_lin_map}}
\label{sec:red_pseudoKrylov_LCLM}

For simplicity, we first assume $s=2$ as the
approach will be easily generalised for arbitrary~$s$.
We follow the standard approach for computing an LCLM
(see e.g.,~\cite[Sec.~4.2.2]{BoChSaLi2012} or~\cite[Algo.~4.27]{Kauers23}).
Let $r_1$, $r_2$ be the respective orders of $L_1$ and $L_2$ and
$d_1$, $d_2$ their respective degrees.  Let
$\alpha= \alpha_1 + \alpha_2$ be the sum of generic solutions of $L_1$
and $L_2$ respectively.  The coefficients of $L$ are read off from a
linear relation between the successive derivatives of $\alpha$.  These
derivatives all lie in the finite dimensional~$k(x)$-vector space spanned by
$A = (\alpha_1, \dots \alpha_1^{(r_1-1)},\alpha_2, \dots,
\alpha_2^{(r_2-1)})$.  Let~$L_1 = p_{1,r_1}\DP_x^{r_1} + \cdots + p_{1,0}$
and
$L_2 = p_{2,r_2}\DP_x^{r_2} + \cdots + p_{2,0}$ with $p_{1,r_1}$ and
$p_{2,r_2}$ nonzero.  By induction suppose that for~$\ell \ge 0$, we
have written
$\alpha^{(\ell)} = \sum_{j=0}^{r_1 - 1 } a_j \alpha_1^{(j)} +
\sum_{j=0}^{r_2 -1 } b_j \alpha_2^{(j)}$ with $a_j,b_j \in k(x)$.
Then differentiation yields
\begin{align*}
  \alpha^{(\ell+1)} = \sum_{j=0}^{r_1 -1} \left( a_j' \alpha_1^{(j)} + a_j \alpha_1^{(j+1)} \right) +
  \sum_{j=0}^{r_2 -1} \left( b_j' \alpha_2^{(j)} + b_j \alpha_2^{(j+1)} \right).
\end{align*}
Finally, we use $L_1(\alpha_1) = L_2(\alpha_2) = 0$ to rewrite
$\alpha_1^{(r_1)}$ and $\alpha_2^{(r_2)}$ on the generating set~$A$.
So if~$V_\ell$ denotes the vector of coefficients of $\alpha^{(\ell)}$
in $A$ for all $\ell$, then
$V_{\ell+1} = (\DP_x+ T )( V_\ell)$,
where~$T \in k(x)^{(r_1 + r_2) \times (r_1 + r_2)}$ is the block diagonal
matrix $\diag(C_1,C_2)$ with $C_i$ the companion matrix associated to
$L_i$, \ie $C_i \in k(x)^{r_i \times r_i}$ is companion with last
column $ -1/p_{i,r_i} \cdot
\begin{bmatrix}
  p_{i,0} &  \cdots & p_{i,r_i-1}
\end{bmatrix}^t$.

\subsection{McMillan degree}
\label{sec:LCLM_McMillan_deg}

We have the description
$C_i = X_i \cdot M_i^{-1}$ with
$M_i = \diag(1,\dots,1,-p_{i,r_i}) \in k[x]^{r_1 \times r_1}$ and
$X_i$ the companion matrix associated to
$(p_{i,0},\dots, p_{i,r_i-1})$. This yields a description
$T = XM^{-1}$ with~$M = \diag(M_1,M_2)$ and $X = \diag( X_1 , X_2)$.
By Proposition~\ref{prop:mcmillandeg_proper_mobius}, there exists a
transformation $\mu = (\alpha x + \beta)/(\gamma x + \delta)$ with
denominator $u$ such that $T_\mu$ is proper (\ie both $C_{1,\mu}$ and
$C_{2,\mu}$ are proper),  and~$\degMM(T) = \degMM(T_\mu)$.  Let
$D_{i,u} = \diag(1,\dots,1, u^{d_i}) \in k[x]^{r_i \times r_i}$ for~$i=1,2$,
then $X_{i,\mu} D_{i,u} = X_i^*$ and
$M_{i,\mu} D_{i,u}= M_{i}^*$ which are both polynomial matrices.
Therefore, $C_{i,\mu} = X_i^* (M_i^*)^{-1}$ for~$i=1,2$ and
$T_\mu = X^* (M^*)^{-1}$ with $X^* = \diag(X_1^*,X_2^*)$ and
$M^* = \diag(M_1^*,M_2^*)$.  So by~(\ref{eq:mcmillandef_def_proper})
and Proposition~\ref{prop:det_den_realisation}, we get
\begin{align*}
  \degMM(C_i) \le  \deg (\det M_i^*) = \deg(u^{d_i} p_{i,r_i}(\mu) ) = d_i.
\end{align*}
Similarly, we also obtain
\begin{align}
  \label{eq:degM_LCLM}
  \degMM(T) \le d_1 + d_2.
\end{align}

\begin{proof}[Proof of Theorem~\ref{thm:degree_bound_LCLM}]
  Suppose first that $s = 2$.
  Let $L_i = \sum_{j=0}^{r_i} p_{i,j}\DP_x^j$ for $i=1,2$ and
  $p_{i,j} \in k[x]$.  The coefficients of $L$ can be directly derived
  from the solution to Problem~\ref{pb:lin_rel_pseudo_lin_map} with
  input $(\DP_x + T,a)$, where $T$ is defined as in
  Section~\ref{sec:red_pseudoKrylov_LCLM} and
  \begin{align*}
    a  = (
    \overbrace{1,0,\dots,0}^{r_1},\overbrace{1,0,\dots,0}^{r_2}
    )^t.
  \end{align*}
  We conclude by applying Theorem~\ref{thm:degree_bound_pseudo_krylov} with
  $d_a = 0$, and $\degMM(T) \le 2d$.

  This result generalises for $s \ge 2$ differential equations with a
matrix $T$ with $s$ different companion blocks on its diagonal. Let
$r_i$ denote the order of $L_i$ for $1 \le i \le s$, $r = \max_i r_i$,
$R = \sum_{i=1}^s r_i$, and $d$ be a bound on the degrees in $x$ of
all the $L_i$'s.  A least common left multiple
$L = \LCLM(L_1,\dots, L_s)$ can be computed as the solution of
Problem~\ref{pb:lin_rel_pseudo_lin_map} for
$T = XM^{-1} \in k(x)^{R \times R}$ and
\begin{align*}
  a=
  (
  \overbrace{1,0,\dots,0}^{r_1},\dots,\overbrace{1,0,\dots,0}^{r_s}
  )^t,
\end{align*}
where $X,M$ are polynomial matrices. As above, one can show that
$\degMM(T) \le sd$.  Hence, the order~$\rho$ of $L$ is less than~$R$
and we conclude in the same way by applying Theorem~\ref{thm:degree_bound_pseudo_krylov} with
  $d_a = 0$, and $\degMM(T) \le s d$.
\end{proof}

\section{Symmetric Product}
\label{sec:sym_prod}

After sums come the closure by products.  In
terms of operators, it corresponds to symmetric product.  For
$L_1,\dots,L_s \in k[x]\left<\DP_x\right>$, we want to compute a
minimal order operator $L = L_1 \otimes \cdots \otimes L_s$ that
annihilates all the products of solutions of the $L_i$'s. We again
first consider the case $s=2$ and then extend the approach for
arbitrary $s$. For $s=2$, we prove the following result.

\begin{theorem}
  \label{thm:deg_bound_symprod}
  Let $L_1, L_2 \in k[x]\left<\DP_x\right>$ be of respective orders
  $r_1$ and $r_2$ and degrees $d_1$ and $d_2$. Then, there exists a
  symmetric product $L= L_1 \otimes L_2$ of order $\rho \le r_1 r_2$
  and degree at most $\rho(d_1 r_2 + d_2 r_1 )$.  For all
  $m \ge \rho$, there exists a left multiple of $L$ of order $m$ and
  degree at most $m(d_1 r_2 + d_2 r_1 )/(m-\rho+1)$.
\end{theorem}

Our new bound is an improvement of the previous best known bound
$r_1^2r_2^2(d_1+d_2)$~\cite[Thm.~8]{Kauers2014}
by one order of magnitude. Experiments suggest that the actual degree
of $L$ is bounded by
$(r_1 r_2 - r_1 -r_2 + 2)(d_1r_2 + d_2 r_1)$~\cite[Chap.~10]{bostan2003these},
and this experimental bound shares the same leading term as our bound.

\subsection{Reduction to Problem~\ref{pb:lin_rel_pseudo_lin_map}}
\label{sec:red_pb1_symprod}

The approach is similar to that of in Section~\ref{sec:red_pseudoKrylov_LCLM}. Let
$r_i$ be the order of $L_i$ and $d_i$ its degree for $i=1,2$. 
Write $L_i = \sum_{j=0}^{r_i}p_{i,j}\DP_x^{j}$ with $p_{i,j} \in k[x]$
and a generic solution $\alpha_i$.  By Leibniz rule, the
successive derivatives of $\alpha = \alpha_1 \alpha_2$ can be written
as $k(x)$-linear combinations of
the elements of~$B = (\alpha_1^{(h)}\alpha_2^{(p)})$ for
$0 \le h < r_1$ and $0 \le p < r_2$. Let
$b_{h,p} = \alpha_1^{(h)}\alpha_2^{(p)} $ and suppose that
\begin{align*}
  \alpha^{(\ell)} = \sum_{h=0}^{r_1-1}  \sum_{p=0}^{r_2-1} e_{h,p}(x) b_{h,p},
\end{align*}
with $e_{h,p} \in k(x)$. We denote by $V_\ell \in k(x)^{r_1 r_2}$ the
vector of coefficients of $\alpha^{(\ell)}$ in this decomposition.
Differentiating gives
\begin{align*}
  \alpha^{(\ell+1)} = \sum_{h,p}\left( e_{h,p}' b_{h,p} +  e_{h,p}(b_{h+1,p} + b_{h,p+1})\right).
\end{align*}
Next the relations $L_1(\alpha_1) = L_2(\alpha_2) = 0$ are used to
rewrite $b_{r_1,p}$ and $b_{h,r_2}$ in the expression above.  Thus,
one can write $V_{\ell+1} = (\DP_x + T) ( V_\ell)$ where $T$ is the
following 
$k(x)$-linear map. It maps any~$b_{h,p}$ for $0 \le h < r_1$ and
$0 \le p < r_2$ to $b_{h+1,p} + b_{h,p+1}$ rewritten according to
$(b_{h,p})$ for $0 \le h < r_1$ and~$0 \le p < r_2$ using $L_1$ and
$L_2$. In the basis
$(b_{0,0},\dots,b_{0,r_2-1},b_{1,0},\dots,b_{r_1-1,r_2-1})$, $T$ can
be seen as the matrix
\begin{align}
  \label{eq:T_matrix_symprod}
  T =  C_1 \otimes I_{r_2} + I_{r_1} \otimes  C_2 \in k(x)^{r_1r_2 \times r_1 r_2},
\end{align}
where $C_i$ denotes the same companion matrix as in
Section~\ref{sec:lclm} and $\otimes$ the Kronecker product.  Also, the
coefficients of $L$ can be directly read off from the solution to
Problem~\ref{pb:lin_rel_pseudo_lin_map} with input $(T,e_1)$
where~$e_1 = (1,0,\dots) \in k(x)^{r_1r_2}$. 

\subsection{McMillan degree}
\label{sec:McMillan_deg_sym_prod}

It remains to bound $\degMM(T)$. Again, let $\mu$ be such that $T_\mu$
is proper (\ie both $C_{1,\mu}$ and $C_{2,\mu}$ are proper).  By
Section~\ref{sec:red_pseudoKrylov_LCLM}, we have the description
$C_{1,\mu} = X_1^*(M_{1}^*)^{-1}$, and $\deg \det M_1^* = d_1$.
Therefore, we have the description
\begin{align*}
  C_{1,\mu} \otimes I_{r_2} = (X_1^* \otimes I_{r_2})\cdot(M_{1}^* \otimes I_{r_2})^{-1},
\end{align*}
and since $C_{1,\mu}$ is proper, so is $C_{1,\mu} \otimes I_{r_2}$, thus
$\degMM(C_1 \otimes I_{r_2}) \le \deg(\det(M_{1}^* \otimes I_{r_2})) =
d_1r_2$.  A similar reasoning shows that
$\degMM( I_{r_1} \otimes C_2 ) \le d_2r_1$, and by
Proposition~\ref{prop:mcmillandeg_prod_inv},
\begin{align}
  \label{eq:degM_symprod}
  \degMM(T) \le  d_1 r_2 + d_2 r_1.
\end{align}

Finally, Theorem~\ref{thm:deg_bound_symprod} follows from
Theorem~\ref{thm:degree_bound_pseudo_krylov} with $d_a$ and $\degMM(T) \le d_1r_2 + d_2r_1$.

\subsection{Symmetric product of several operators}
\label{sec:several_operators_symprod}

This can be generalised to the computation of a symmetric product of
$s \ge 2$ operators. Let~$L = L_1 \otimes \cdots \otimes L_s$ with
$L_i$ of order $r_i$ and degree at most $d$ for all $i$.  Let
$R = r_1 \cdots r_s$ and $r = \max_i(r_i)$.
Let~$\alpha = \alpha_1 \cdots \alpha_s$ where $L_i(\alpha_i) = 0$ for all
$i$.  By Leibniz rule, one can express the successive derivatives
of~$\alpha$ as linear combinations of
$(\alpha_1^{(h_1)}\cdots \alpha_s^{(h_s)})$ for $0 \le h_i < r_i$,
$1 \le i \le s$.  Therefore~$L$ can still be derived from a solution
of Problem~\ref{pb:lin_rel_pseudo_lin_map} with
$T \in k(x)^{R \times R}$. And one can show
that~$\degMM(T) \le \sum_{i=1}^s (r_1 \cdots r_{i-1} \cdot d \cdot
r_{i+1}\cdots r_s)$, generalising the proof
of~(\ref{eq:degM_symprod}).  Then, $\degMM(T) \le sdr^{s-1}$ so $L$ is
of order at most $R$ and degree bounded by
$sRdr^{s-1} \le sdr^{2s-1}$. By comparing with the previous
best known bound
$sR^2d \le sdr^{2s}$~\cite[Thm.~8]{Kauers2014}, we observe that
we save the same factor $r$ as in the case~$s=2$.

\section{Polynomials}
\label{sec:closure_polynomials}

In Sections~\ref{sec:lclm} and~\ref{sec:sym_prod}, we gave bounds for
the effective closure properties of D-finite functions by sums and
products.  More generally, any function $\alpha$ that can be expressed
as a polynomial in some D-finite functions $\alpha_1,\dots,\alpha_s$
and their derivatives is also D-finite~\cite{stanley1980dfinite}.  In
this section, we
bound the arithmetic size of an annihilator of
$\alpha$ in terms of the size of the annihilators of the $\alpha_i$'s
and improve the previous bound~\cite{Kauers2014}.

Suppose that $\alpha_1, \dots, \alpha_s$ are generic solutions of the
operators $L_1,\dots,L_s$ with orders $r_1,\dots,r_s$ and degree
$d_1,\dots, d_s$ respectively, we want an annihilator $L$ of
\begin{align}
  \label{eq:poltodiffeq_alpha}
  \alpha = J(x,\alpha_1,\dots, \alpha_1^{(r_1-1)},\dots, \alpha_s,\dots, \alpha_s^{(r_s -1 )}),
\end{align}
where $J(x,y_{1,0},\dots,y_{1,r_1-1},\dots,y_{s,0},\dots,y_{s,r_s-1})$
is a multivariate polynomial. To derive degree bounds for $L$, one
could decompose $J$ as sums and products and obtain bounds from
Theorem~\ref{thm:degree_bound_LCLM} and~\ref{thm:deg_bound_symprod}.
But this direct method
fails in establishing tight degree bounds~\cite{Kauers2014} as it
does not exploit the structure of the polynomial $J$.
Instead, we adopt the settings of~\cite{Kauers2014}:
the operator $L$ is computed 
directly from the polynomial $J$ by expressing the successive
derivatives of $\alpha$ until a linear relation is found.  We denote
$\deg_x(J)$ for the degree of $J$ with respect to $x$ and
$\Deg(J) =(k_1,\dots,k_s)$ where~$k_i$ is the total degree of $J$ in
the variables $(y_{i,j})_{0\le j < r_i}$. We further assume that $J$
is homogeneous with respect to each group of variables
$(y_{i,j})_{0\le j < r_i}$ and of degree $k_i$ for $1\le i\le s$.
We further refer to this assumption by saying that $J$ is \emph{homogeneous}.
If $J$ is not
homogeneous, then one can decompose~$J = J_1 + \cdots + J_\ell$ where
every $J_i$ is homogeneous, get a bound for each $J_i$ and combine
them using Theorem~\ref{thm:degree_bound_LCLM}. It turns out that the
previous overestimation does not occur when every homogeneous part is
handled as a whole~\cite{Kauers2014}.

\begin{theorem}
  \label{thm:poltodiffeq_bound}
  Let $L_1,\dots,L_s \in k[x]\left<\DP_x\right>$ be of respective
  orders $r_i$ and degrees $d_i$ for $1\le i \le s$.  Let $J$ be a
  homogeneous multivariate polynomial as
  in~(\ref{eq:poltodiffeq_alpha}) with $\Deg(J) = (k_1, \dots,
  k_s)$. Then, there exists an operator $L \in k[x]\left<\DP_x\right>$
  such that for all $\alpha_1,\dots,\alpha_s$ such that
  $L_1(\alpha_1) = \cdots = L_s(\alpha_s) = 0$ and
  $\alpha = J(x,\alpha_1, \dots, \alpha^{(r_1-1)},\dots, \alpha_s,\dots,
  \alpha_s^{(s -1 )})$, we have $L(\alpha) = 0$ and $L$ has order
  $\rho\le R = \prod_{i=1}^s \binom{k_i+r_i-1}{k_i}$
  and degree at most
  $\rho \deg_x(J) + \rho R \sum_{i=1}^s\frac{k_id_i}{k_i +
    r_i -1}$.  For all $m \ge \rho$, there exists a nonzero operator
  of order $m$, which also annihilates the function $\alpha$, of
  degree at most
  $1/(m-\rho+1)\cdot(\rho \deg_x(J) + m \cdot R
  \sum_{i=1}^s\frac{k_id_i}{k_i + r_i -1})$.
\end{theorem}

This result improves the previous degree bound
$R\deg_x(J) + R^2
\sum_{i=1}^sk_id_i$~\cite[Thm.~8]{Kauers2014}
for an operator of minimal order. Similarly it improves the
associated order-degree
curves~\cite[Thm.~13]{Kauers2014}.  Note that if
one applies this result in the case of the symmetric product (\ie with
$J = y_{1,0}y_{2,0}$), one retrieves
Theorem~\ref{thm:deg_bound_symprod} as a corollary.
Other special cases of this general
theorem are presented in Section~\ref{sec:applications_poltodiffeq}.

\subsection{Reduction to Problem~\ref{pb:lin_rel_pseudo_lin_map}}
\label{sec:red_pb1_poltodiffeq}

Using Leibniz's rule, the successive derivatives of $\alpha$ all lie
in the $k(x)$-vector space $\mathcal V$ spanned by the elements
\begin{align}
  \label{eq:elements_gen_set_poltodiffeq}
  \left(\DP_x^{h_{1,1}}\alpha_1\right)\cdots \left(\DP_x^{h_{1,k_1}}\alpha_1\right)\cdots
  \left(\DP_x^{h_{s,1}}\alpha_s\right)\cdots \left(\DP_x^{h_{s,k_s}}\alpha_s\right),
\end{align}
with $0 \le h_{i,1} \le \cdots \le h_{i,k_i} < r_i$ for all
$1 \le i \le s$. The dimension of $\mathcal V$ is at most
$R = \prod_{i=1}^s \binom{k_i+r_i-1}{k_i}$ and so is the
order of $L$.  Now, if $V_\ell$ is the vector of coefficients of
$\alpha^{(\ell)}$ in the above decomposition, we have
$V_{\ell +1} = \theta V_\ell$ where $\theta$ is the pseudo-linear map
$ \DP_x +T$ with $T \in k(x)^{R \times R}$ mapping
any above element to its derivative where we rewrite high order
derivatives using $L_1, \dots, L_s$. Moreover, one can read $V_0$ from
$\alpha$ given by~(\ref{eq:poltodiffeq_alpha}). Hence, one can compute
the coefficients of $L$ from a solution to
Problem~\ref{pb:lin_rel_pseudo_lin_map} for $(T,V_0)$.

\subsection{McMillan degree}
\label{sec:McMillan_poltodiffeq}

The columns of $T$ consist of the derivatives of the elements of the
generating set as in~(\ref{eq:elements_gen_set_poltodiffeq}) according
to this same generating set. To each column of $T$ corresponds a
tuple
\begin{align*}
  \mathbf h = (h_{1,1},\dots,h_{1,k_1},\dots,h_{s,1},\dots,h_{s,k_s}),
\end{align*}
with $0 \le h_{i,1} \le \cdots \le h_{i,k_i} < r_i$ for all
$1 \le i \le s$.  We denote by $\mbf H$ the set of columns of $T$.
Generalising the approach of Section~\ref{sec:red_pseudoKrylov_LCLM}, we write
$T= XM^{-1}$ where $M \in k[x]^{R \times R}$ is a diagonal matrix
whose diagonal entries are the denominators of the columns of $T$, and
$X = TM \in k[x]^{R \times R}$ is a polynomial
matrix.  Let $q_i(x)$ be the leading coefficient of $L_i$ (\ie the
coefficient of $\DP^{r_i}$).
Differentiating~(\ref{eq:elements_gen_set_poltodiffeq}) and rewriting
high order derivatives using $L_1, \dots, L_s$ show that the
derivative of~(\ref{eq:elements_gen_set_poltodiffeq}) can be written
with denominator
\begin{align*}
  q_{\mbf h} = \prod_{i=1}^s q_i^{\delta({h_{i,k_i},r_i-1})},  
\end{align*}
where
$\delta({a,b}) =1$ if $a=b$, $0$ otherwise.~These products give the
diagonal entries of $M$.~It has degree bounded by
$d_{\mbf h} = \sum_{i=1}^s \delta({h_{i,k_i},r_i-1})\cdot d_i$ and the
corresponding column of $X$ satisfies the same degree bound.  Now, for
$\mu$ a transformation such that $T_\mu$ is proper with denominator
$u$, we define $D_u = \diag(u^{d_{\mbf h}})_{\mbf{h} \in \mbf H}$.
Then, $X^* = X_\mu D_u$ and $M^* = M_\mu D_u$ are both polynomial
matrices whose column $\mbf{h}$ has degree bounded by $d_{\mbf{h}}$
for all $\mbf h \in \mbf H$.  So, $T_\mu = X^* (M^*)^{-1}$ and is
proper. So,
$\degMM(T) = \deg(\det M^*) \le \sum_{\mbf h \in \mbf H} d_{\mbf h} =
\sum_{i=1}^s R_i d_i$, where $R_i$ is the number
of elements in $\mbf{H}$ satisfying $h_{i,k_i} = r_{i}-1$.  Thus, we have
\begin{align*}
  R_i = R
  \frac{\binom{k_i + r_i -2}{k_i-1}}{\binom{k_i + r_i -1}{k_i}}
  = \frac{R k_i}{k_i + r_i - 1}.  
\end{align*}
Finally, it follows that
\begin{align}
  \label{eq:degMMT_polynomials}
  \degMM(T) \le R \sum_{i=1}^s d_i \frac{k_i}{k_i + r_i -1}.
\end{align}

Again, Theorem~\ref{thm:poltodiffeq_bound} is obtained as a consequence of
Theorem~\ref{thm:degree_bound_pseudo_krylov}, with $d_a = \deg_x(J)$
and~(\ref{eq:degMMT_polynomials}).

\subsection{Applications}
\label{sec:applications_poltodiffeq}

We already mention the symmetric product
as a consequence of Theorem~\ref{thm:poltodiffeq_bound}.
We exemplify
Theorem~\ref{thm:poltodiffeq_bound} through three other special cases:
symmetric power, associate and Wronskian. For a family of power series
$(\alpha_1, \dots, \alpha_r)$, we define its Wronskian
$w(\alpha_1, \dots, \alpha_r)$ as the determinant
of~$(\alpha_j^{(i-1)})_{1\le i,j\le r}$. 
\begin{corollary}
  \label{cor:appli_poltodiffeq}
  \begin{enumerate}
  \item (Symmetric Power) Let $L \in k[x]\left<\DP_x\right>$ of order
    $r$ and degree $d$.  Then, there exists an operator
    $M \in k[x]\left<\DP_x\right>$ such that for each solution $\alpha$
    of $L$, we have $M(\alpha^\ell) = 0$ and~$M$ has order at most
    $\rho = \binom{\ell+r-1}{\ell}$ and degree at most $\rho^2 \frac{\ell d}{\ell +r-1}$.
  \item (Associate) Let $L \in k[x]\left<\DP_x\right>$ of order $r$
    and degree $d$ and another operator $A \in k[x]\left<\DP_x\right>$
    of order $<r$ and degree $d_A$.  Then, there exists an operator
    $M \in k[x]\left<\DP_x\right>$ such that for each solution $\alpha$
    of $L$, we have $M\cdot A(\alpha) = 0$ and $M$ has order at most
    $r$ and degree at most $r(d_A + d)$.
  \item (Wronskian) Let $L_1,\dots,L_r \in k[x]\left<\DP_x\right>$ all
    of order $r$ and degree $d$. Then, there exists an operator
    $M \in k[x]\left<\DP_x\right>$ such that for all
    $\alpha_1,\dots,\alpha_r$ such that
    $L_1(\alpha_1) = \cdots = L_r(\alpha_r) = 0$, we have
    $M(w(\alpha_1,\dots,\alpha_r)) = 0$ and $M$ has order at most
    $r^r$ and degree at most $r^{2r}d$.
  \end{enumerate}
\end{corollary}

\begin{proof}
  Part 1~is Theorem~\ref{thm:poltodiffeq_bound} with $J = y_{1,0}^\ell$,
  and thus $s=1$, $\Deg(J) = \ell$ and $\deg_x(J) = 0$. Similarly, in Part 2,
  if $A = \sum_{j=0}^{r-1} a_j(x)\DP_x^j$, then we have
  $J = \sum_{j=0}^{r-1} a_j y_{1,j}$, and thus $s=1$, $\Deg(J) =1$
  and~$\deg_x(J) = d_A$ and the result follows. Finally, in Part 3, we
  have $J = \det (y_{i,j-1})_{1\le i,j \le r}$, and thus~$s=r$,
  $\Deg(J) = (1,\dots,1)$ and $\deg_x(J) = 0$ and the result again
  follows from Theorem~\ref{thm:poltodiffeq_bound}.
\end{proof}

Each part of Corollary~\ref{cor:appli_poltodiffeq} benefits from the
refined bound in Theorem~\ref{thm:poltodiffeq_bound} to 
improve the previously best known bound~(cf.~\cite{Kauers2014}).
The special case of symmetric powers is relevant in many applications.
Indeed, several algorithms for solving linear differential equations rely
on the computation of symmetric powers~(see e.g.,~\cite[Sec.~4.3.3]{vdPuSi2012galois}).
In particular for differential operators of order 2, there is a fast
algorithm for computing symmetric powers in this specific
situation~\cite{Chalkley1989,BrMuWe1997}. 
This algorithm exploits the
structure of the involved pseudo-Krylov matrix and avoids solving the
linear system.

\section{Differential equation for algebraic function}
\label{sec:algeqtodiffeq}

Let $P(x,y) \in k[x,y]$ be a nonzero bivariate polynomial.
All the roots $\alpha(x)$ of $P$ are D-finite~\cite{abeloeuvres} and  
we address the problem of
computing a nonzero (minimal order) differential operator annihilating all these roots.
Cockle's
algorithm~\cite{cockle1861,BoChSaLeSc2007}
computes such an operator as a linear relation between the
images of $\alpha^{(i)}$ modulo $P$. In fact, this algorithm solves
an instance
of Problem~\ref{pb:lin_rel_pseudo_lin_map}~\cite[Sec.~2.1]{BoChSaLeSc2007}.
Here we make this particular instance explicit, and then derive the following
degree bounds.

\begin{theorem}
  \label{thm:degbd_algeqtodiffeq}
  Let $P \in k[x,y]$ be square-free with respect to $y$,
  $(d,r) = (\deg_x(P),\deg_y(P))$.  Then, there exists a differential
  resolvent of $P$ of order $\rho \le r$ and degree at most
  $\rho (2r-1)d$.  Moreover, for all $m \ge \rho$, there exists a
  nonzero operator of order $m$ annihilating all the roots of $P$ of
  degree at most $m/(m-\rho+1)\cdot (2r-1)d$.
\end{theorem}

For a differential resolvent, our bound is in $O(r^2d)$ and governed by
$2r^2d$.  The previous best known bound behaves like
$4r^2d$~\cite[Thm.~1]{BoChSaLeSc2007} and its proof is \emph{ad hoc},
namely, not derived from the analysis of the algorithm.
So this is an improvement of the
constant from 4 to 2.

Experimentally, if we assume that $r = d$, we observe that the output
degree is bounded by~$d(2d^2 - 3d +3)$. Our bound gives $d(2d^2 - d)$
and the previous bound is
$d(4d^2 -11d/2 + 7/2)$~\cite{BoChSaLeSc2007}.  This is the
first time that a bound with the first term $2d^3$ is obtained.

Let $P(x,y) \in k[x,y]\setminus \{0\}$ be square-free with respect to
$y$ and let $d$ be its degree in $x$ and $r$ in~$y$.  We write
$P_x = \DP_x(P)$ and $P_y = \DP_y(P)$. In this section, $\mathcal A$
denotes the vector space $k(x)[y]_{< r}$.

\subsection{Reduction to Problem~\ref{pb:lin_rel_pseudo_lin_map}}
\label{sec:algeqtodiffeq_red_pb1}

Let $\alpha_1, \dots, \alpha_{r}$ be the roots of $P$ in an algebraic
closure of $k(x)$.  For any root $\alpha$ of $P$, by differentiating
$P(x,\alpha) =0$ and since $P$ is square-free, we have
$\DP_x (\alpha) = - P_x(x,\alpha) / P_y(x,\alpha)$.  Hence the
derivation $\DP_x$ uniquely extends to
$k(x)(\alpha_1,\dots,\alpha_{r})$.  By induction, for all $i\ge1$,
there exists a polynomial $C_i \in k[x,y]$ such that
$\alpha^{(i)} = C_i(x,\alpha) / P_y(x,\alpha)^{2i-1}$, for any root
$\alpha$ of $P$.  Since $P$ and~$P_y$ are coprime, one can define
$D_i(x,y)$ to be the polynomial in $\mathcal A$
such that $D_i(x,\alpha) = \alpha^{(i)}$ for each root~$\alpha$ of
$P$. Note that $D_0 = y$.  And thus it suffices to find the minimal
linear dependence between the $D_i$'s to compute the differential
resolvent of $P$.

Moreover, by differentiating $\alpha^{(i)} = D_i(x,\alpha)$, we obtain
\begin{align*}
  \alpha^{(i+1)} &= \DP_x (D_i(x,\alpha)) + \DP_x( \alpha) \cdot \DP_y (D_i(x,\alpha)) \\
                 &= \DP_x (D_i(x,\alpha)) - \frac{P_x(x,\alpha)}{P_y(x,\alpha)} \DP_y (D_i(x,\alpha)).
\end{align*}
Therefore $D_{i+1} = (\DP_x + T)( D_i)$, where $T$ is the
$k(x)$-linear map
$T \colon a \mapsto - \DP_y(a) P_x /P_y ~\mathrm{mod}~P$ on
$\mathcal A$. Finally, the differential resolvent of $P$ is the
solution of Problem~\ref{pb:lin_rel_pseudo_lin_map} for $(T,y)$.

\subsection{McMillan degree}
\label{sec:McMillan_algeqtodiffeq}
Fix the monomial basis $(1,y,\dots,y^{r-1})$ of $\mathcal A$ and view
$T$ as a matrix in $k(x)^{r \times r}$. 

\begin{lemma}
  \label{lem:algeqtodiffeq_XMm1Y}
  There exist matrices
  $X \in k^{r \times (2r-1)}, M \in k[x]^{(2 r-1) \times (2r-1)}$ and
  $Y \in k[x]^{(2 r-1) \times r}$, with~$M$ invertible, $\deg M \le d$
  and $\deg Y \le d-1$, such that $T = XM^{-1}Y$.
\end{lemma}

\begin{proof}
  For $a \in \mathcal A$, $T a$ is the solution $V$ of the following
  B{\'e}zout equation
  \begin{align*}
    -\DP_y(a)P_x = U \cdot P + V \cdot P_y.
  \end{align*}
  So if
  $Y \colon a \in \mathcal A \mapsto -\DP_y(a)P_x \in k(x)[y]_{< 2r
    -1}$ and $X$ is the projection $(U,V) \mapsto V$ onto the last
  $r$ coordinates, we have $T = X M^{-1}Y$ where $M$ is the
  Sylvester map associated to $P$ and $P_y$ which is invertible since
  $P$ is square-free, and $X,M,Y$ satisfy the required property.
\end{proof}

By Proposition~\ref{prop:mcmillandeg_proper_mobius}, let $\mu$ be such
that $T_\mu$ is proper and $\degMM(T_\mu) = \degMM(T)$ and denote by
$u$ the denominator of $\mu$.  Then, $T_\mu = X M_\mu^{-1}Y_\mu$ and
$M_\mu = 1/u^d \cdot M^*$, $Y_\mu = 1/u^{d-1} \cdot Y^*$, where
$M^*,Y^*$ are both polynomial matrices with $\deg M^* \le d$ and
$\deg Y^* \le d-1$.  So, $T_\mu = X(M^*)^{-1}\cdot u
Y^*$. By~(\ref{eq:mcmillandef_def_proper}) and
Proposition~\ref{prop:det_den_realisation}, we get
\begin{align}
  \label{eq:degMcMillan_T_algeqtodiffeq}
  \degMM(T) \le \deg (\det M^*) \le (2r-1)d.
\end{align}

Finally, Theorem~\ref{thm:degbd_algeqtodiffeq}
follows from Theorem~\ref{thm:degree_bound_pseudo_krylov}
and~(\ref{eq:degMcMillan_T_algeqtodiffeq}), with $T$ and
$a = (0,1,0,\dots,0)^t \in k[x]^r$ since $D_0 = y$.

\section{Composition of algebraic and D-finite functions}
\label{sec:subs_algDfinite}

It is known that for any algebraic function $g$ and D-finite function
$f$, the composition $h = f \circ g$ is again
D-finite~\cite{stanley1980dfinite}.  The case of the differential
resolvent is the particular case where $f = x$.  We apply the
same approach as in Section~\ref{sec:algeqtodiffeq} by extending
Cockle's algorithm in order to derive degree bounds.  We will compare our
results with the previous known
bounds~\cite{KaPo2017}.

Let $P(x,y) \in k[x,y]$ be a nonzero bivariate polynomial which is
square-free with respect to $y$, and let $d_P$ be its degree in $x$
and $r_P$ in $y$.  Again, $P_x$ (resp.~$P_y$) denotes the
partial derivative of $P$ with respect to $x$ (resp.~$y$).  Let
$L \in k[x]\left<\DP_x\right>$ be a nonzero differential operator of
degree $d_L$ and order $r_L$.  Let $l(x) \in k[x]$ be the leading
coefficient of $L$. We require that $\gcd(P(x,y),l(y)) =1$.  This is
the case when $P$ is assumed to
have no divisor in $\bar k(y)$ where $\bar k$ is an algebraic closure
of $k$~\cite{KaPo2017}.
We want a nonzero operator $M$ annihilating all the functions
$h = f \circ g$ where $L ( f) = 0$ and $g$ is a root of $P$.
Again, the method consists in computing a relation between the
successive derivatives of a generic solution $h = f \circ g$ modulo
the relations $P(x,g) = 0$ and $(L ( f))(g(x)) = 0$.
This enables us to view the problem as an instance of
Problem~\ref{pb:lin_rel_pseudo_lin_map} and prove
the following degree bound.

\begin{theorem}
  \label{thm:deg_bd_subsalgDfinite}
  Let $P \in k[x,y]$ be square-free with respect to $y$,
  $(d_P,r_P) = (\deg_x(P),\deg_y(P))$
  and~$L \in k[x]\left<\DP_x\right>$ with leading coefficient $l \in k[x]$
  such that $l(y)$ is coprime with $P(x,y)$, and let
  $(d_L,r_L) = (\deg_x(L),\deg_{\DP_x}(L))$.  Then, there exists
  $M \in k[x]\left<\DP_x\right> \setminus \{0\}$ annihilating all the
  power series $h = f \circ g$ where $L( f) = 0 $ and
  $P(x,g(x))= 0$, of order $\rho \le r_P r_L$ and degree at most
  $\rho d_P(r_L(2r_P-1) + d_L)$.  Also, for all $m \ge \rho$ there
  exists an operator of order $m$ annihilating all these power series
  of degree at most $m/(m-\rho +1) d_P(r_L(2r_P-1) + d_L)$.
\end{theorem}

For minimal annihilators, our degree bound is in
$O(r_Pr_L d_P (r_L r_P + d_L))$ which is an improvement of the
previous bound in
$O(r_Pr_L d_P (r_Lr_P +
r_Ld_L))$~\cite[Thm.~8]{KaPo2017}.  Moreover, the leading
term of our bound ($d_Pr_Lr_P(2r_Lr_P + d_L)$) coincides with the one
conjectured from
experiments~\cite[Conj.~10]{KaPo2017}.

\subsection{Reduction to Problem~\ref{pb:lin_rel_pseudo_lin_map}}
\label{sec:comp_alg_Dfinite_red_prob1}

As in Section~\ref{sec:algeqtodiffeq_red_pb1}, we have
$g' = -P_x(x,g)/P_y(x,g)$ for any root $g$ of $P$. Thus,
\begin{align*}
  h' = (f\circ g) = g' \cdot(f' \circ g) = -\frac{P_x(x,g)}{P_y(x,g)} f' \circ g.
\end{align*}
By induction on $\ell \ge 0$, using $g' = -P_x(x,g)/P_y(x,g)$ and
$(L ( f))(g(x)) = 0$, it follows that
\begin{align*}
  h^{(\ell)} = (f \circ g)^{(\ell)} = \frac{1}{U(x,g)^\ell} \sum_{j=0}^{r_L-1}E_{i,j}(x,g)(f^{(j)}\circ g),
\end{align*}
where
$U(x,y) = P_y^2(x,y)l(y)$~\cite[Lem.~4]{KaPo2017}.
Note that $U(x,g) \neq 0$ since $P$ and $l(y)$ are coprime and~$P$ is square-free.  So, one
can define
\begin{align*}
  V_\ell(x,y,\DP) = \sum_{i=0}^{r_L-1}\sum_{j=0}^{r_P-1}e_{i,j}(x)y^j \DP^i,
\end{align*}
with $e_{i,j}(x) \in k(x)$ such that
$h^{(\ell)}(x) = \left(V_\ell(x,g(x),\DP)( f)\right)\left(g(x)\right)$.  We 
reduce the computation to linear algebra problem in the $k(x)$-vector space
$\mathcal A = k(x)[y]_{< r_P}[\DP]_{< r_L}$. According to this
decomposition, we have $V_0 = 1$. Differentiating
\begin{align*}
  h^{(\ell)}(x)  =  \sum_{i=0}^{r_L-1}\sum_{j=0}^{r_P-1}e_{i,j}(x)g(x)^j f^{(i)}(g(x)),
\end{align*}
we obtain
\begin{align*}
  h^{(\ell + 1)}(x)  =  \sum_{i=0}^{r_L-1}\sum_{j=0}^{r_P-1}\left( e'_{i,j}(x)g(x)^j f^{(i)}(g(x))
  + e_{i,j}(x)g'\left(j g^{j-1}(x)f^{(i)}(g(x)) + g^{j}(x)f^{(i+1)}(g(x)) \right)\right).
\end{align*}
Thus, it follows that
\begin{align*}
  V_{\ell +1} = \DP_x V_\ell + T \cdot V_\ell = \theta V_\ell,
\end{align*}
with $T$ the following $k(x)$-linear map on $\mathcal A$
\begin{align*}
  T \colon a \in \mathcal A \mapsto - \frac{P_x}{P_y}\left( \DP_y(a) + a \cdot \DP \right)
  ~ \text{ mod } \left< L(y,\DP), P(x,y)\right>.
\end{align*}
Finally, one can compute the operator $M$ from a solution of
Problem~\ref{pb:lin_rel_pseudo_lin_map} for $(T,V_0)$.

\subsection{McMillan degree}
\label{sec:compalgDfin_McMillan}

Fix the basis
$B = (1,y,\dots,y^{r_P-1},\DP, y \DP, \dots,
y^{r_P-1}\DP,\dots,y^{r_P-1}\DP^{r_L-1}) = (y^{i}\DP^{j})_{1\le j<
  r_L, 1 \le i < r_P}$ of~$\mathcal A$ and view $T$ as a matrix in
$k(x)^{r_Pr_L \times r_P r_L}$.

First, one can decompose $T = T_1 T_2$ where
$T_1 \colon a \in \mathcal A \mapsto \left(-P_x/P_y \cdot a \text{ mod
  } P\right) \in \mathcal A$ and
$T_2 \colon a \in \mathcal A \mapsto \DP_y(a) + \left(a \cdot \DP
  \text{ mod } \left<L(y,\DP),P(x,y)\right>\right) \in \mathcal A$.
By Proposition~\ref{prop:mcmillandeg_prod_inv},
\begin{align*}
  \degMM(T) \le \degMM(T_1) + \degMM(T_2).
\end{align*}
We bound $\degMM(T)$ by bounding the McMillan
degrees of $T_1$ and $T_2$ separately.

Next, we have $T_1 = I_{r_L} \otimes T_{\mathrm{alg}}$ where
$T_{\mathrm{alg}}$ is the matrix for the differential resolvent in
Section~\ref{sec:algeqtodiffeq_red_pb1}.  Thus
by~(\ref{eq:degMcMillan_T_algeqtodiffeq}),
$\degMM(T_{\mathrm{alg}}) \le (2r_P - 1)d_P$, and using an irreducible
description of $T_{\mathrm{alg}}(\mu(x))$ for $\mu$ of the form of
Proposition~\ref{prop:mcmillandeg_proper_mobius}, a similar argument
to that of Section~\ref{sec:McMillan_deg_sym_prod} shows that 
\begin{align*}
  \degMM(T_1) = \degMM( I_{r_L} \otimes T_{\mathrm{alg}}) \le r_L(2r_P
- 1)d_P.
\end{align*}

For the matrix $T_2$, one can first notice that the map
$\mathrm{diff}_y \colon a \in \mathcal \mapsto \DP_y(a) \in \mathcal
A$ corresponds to a matrix with scalar coefficients, so its McMillan
degree is 0.  Therefore, $\degMM(T_2) = \degMM(T_\DP)$ where~$T_\DP$ is
the matrix of the multiplication by $\DP$ (on the right) in
$\mathcal A$.  We investigate the matrix $T_\DP$ of the multiplication
by $\DP$ in $\mathcal A$ written in the basis $B$. Let us write
$L(y,\DP) = l(y)\DP^{r_L} + \sum_{i=0}^{r_L-1}l_{i}(y)\DP^i$.  From
the identity
\begin{align*}
  y^i\cdot \DP^{r_L} = \sum_{j=0}^{r_L -1 } -\frac{l_j(y)}{l(y)}y^i \DP^j \mod
  \left< L(y,\DP),P(x,y)\right>, 
\end{align*}
we obtain that $T_\DP$ has the following block decomposition:
\begin{align}
  \label{eq:T_DP_block_decomposition}
  T_\DP =
  \begin{bmatrix}
    0_{r_P} &&& S_0 \\
    I_{r_P} & \ddots& &S_1\\
            & \ddots & & \vdots\\
            & &I_{r_P} & S_{r_L-1}
  \end{bmatrix} \in k(x)^{r_Pr_L \times r_P r_L},
\end{align}
where $S_i \in k(x)^{r_P \times r_P}$ is the matrix of the
multiplication by $-l_i(y)/l(y)$ in $k(x)[y]/(P)$ in the
basis~$(1,y,\dots,y^{r_P-1})$ (recall that $l(y)$ is invertible mod $P$
since $\gcd(P,l(y)) = 1$).  Moreover, we observe that for any
transformation $\mu(x)$ of the form of
Proposition~\ref{prop:mcmillandeg_proper_mobius} with denominator $u$,
$S_i(\mu(x))$ is the matrix of the multiplication by $-l_i/l$ modulo
$P(\mu(x),y)$ which is equal to the matrix of the multiplication by
$-l_i/l$ modulo $P^*(x,y) \in k[x,y]$ where
$P^*(x,y) =P(\mu(x),y) \cdot u^{d_P}$ has degree in $x$ at most
$d_P$. Therefore, up to replacing $P$ by $P^*$, one can assume that
$T_\DP$ is proper. Thus, it suffices to find a description of $T_\DP$
and bound the degree of the determinant of the denominator.
We prove the following result for $T_\DP$ in Section~\ref{sec:pf_desc_T_DP}.
\begin{proposition}
  \label{prop:description_T_DP}
  There exists a description $T_\DP = XM^{-1}$
  with $X,M \in k[x]^{r_Pr_L \times r_P r_L}$ and $\deg \det M \le d_L d_P$.
\end{proposition}

\noindent Therefore, it follows that
\begin{align*}
  \degMM(T_2) = \degMM(T_\DP) = \deg \det M \le d_Ld_P,
\end{align*}
and finally, we conclude:
\begin{align}
  \label{eq:degMM_T_subsalgDfinite}
  \degMM(T) \le \degMM(T_1) + \degMM(T_2) \le d_P(r_L(2r_P-1) + d_L).
\end{align}

Applying Theorem~\ref{thm:degree_bound_pseudo_krylov}
with $T$, $a = (1,0,\dots)^t \in k[x]^{r_P r_L}$ (since $V_0 = 1$)
and~(\ref{eq:degMM_T_subsalgDfinite})
completes the proof of Theorem~\ref{thm:deg_bd_subsalgDfinite}.

\subsection{Proof of Proposition~\ref{prop:description_T_DP}}
\label{sec:pf_desc_T_DP}

We exhibit a description for
the matrix $T_\DP$ in~(\ref{eq:T_DP_block_decomposition}).
We proceed as follows: we prove that there exists
a non-singular polynomial matrix
$Q \in k[x]^{r_P \times r_P}$ with $\deg (\det Q) \le d_L d_P$
such that for all~$0 \le i < r_L$,
we have a description
$S_i = N_i Q^{-1}$ with $N_i \in k[x]^{r_P \times r_P}$.
Then, we deduce the description
\begin{align}
  \label{eq:T_DP_description}
  T_\DP =
  \begin{bmatrix}
    0_{r_P} &&& N_{l_0} \\
    I_{r_P} & \ddots& &N_{l_1}\\
            & \ddots & & \vdots\\
            & &I_{r_P} & N_{l_{r_L-1}}
  \end{bmatrix}
  \begin{bmatrix}
    I_{r_P(r_L-1)} & 0 \\ 0 & Q
  \end{bmatrix}^{-1},
\end{align}
that satisfies the requirement of Proposition~\ref{prop:description_T_DP}.

For a polynomial $g \in k(x)[y]$, we denote by $M_g \in k(x)^{r_P \times r_P}$
the (rational) matrix  of the
multiplication by $g$ modulo $P$ in the monomial basis.
For all
$0 \le i < r_L $, $S_i = M_{l_i}\cdot M_{l}^{-1}$.  We denote by
$p_i \in k[x]$ the coefficients (w.r.t.~$y$) of degree $i$ in $P(x,y)$.
One can prove the following results for matrices of multiplication.

\begin{lemma}
  \label{lem:realisation_powMy}
  Let $M_y \in k(x)^{r_P \times r_P}$ be the matrix of multiplication
  by $y$ modulo $P(x,y)$, \ie the companion matrix associated to $P$.
  For all $\ell \ge 1$, there exist polynomial
  matrices $D_\ell, U_\ell$ with $D_\ell$ of size $\ell r_P \times \ell r_P$ and
  $\det D_\ell = p_{r_P}^\ell \neq 0$, and $U_\ell =
  \begin{bmatrix}
    I_{r_P} & 0
  \end{bmatrix}^t \in k[x]^{\ell r_P \times r_P}$, such that for all $0 \le k \le \ell$,
  there exists $V_k^{(\ell )} \in k[x]^{r_P \times \ell r_P }$ giving the realisation
  \begin{align*}
    M_{y^k} = (M_y)^k = V_k^{(\ell)} D_\ell^{-1}U_\ell.
  \end{align*}
\end{lemma}

The proof is based on a construction to build a realisation of a product
of matrices $R_1 R_2$ from a realisation of $R_1$
and a realisation of $R_2$~\cite[Lem.~3]{coppel1974}.

\begin{proof}
  By induction on $\ell \ge 1$.  For $\ell = 1$, let $U_1 = I_{r_P}$,
  $D_1 = \mathrm{diag}(1,\dots,1,p_{r_P})$ and $V_1^{(1)}$ the
  companion matrix whose last column is $-(p_0,\dots, p_{r_P-1})^t$.
  Then $M_y = V_1^{(1)}D_1^{-1}U_1$ and
  $M_{y^0} = I_d = V_1^{(0)}D_1^{-1}U_1$ with $V_1^{(0)} = D_1$, so
  the result holds.  Suppose the result holds for $\ell -1 \ge 1$. We
  have
  \begin{align*}
    M_{y^\ell} = M_y \cdot M_{y^{\ell -1}} = V_1^{(1)}D_1^{-1}U_1
    \cdot  V_{\ell-1}^{(\ell-1)}D_{\ell-1}^{-1}U_{\ell-1}.
  \end{align*}
  Now define $U_\ell =
  \begin{bmatrix}
    { U_{\ell-1}^t} & 0_{r_P}
  \end{bmatrix}^t \in k[x]^{\ell r_P \times r_P}$,
  \begin{align*}
    D_\ell = \begin{bmatrix}
      D_{\ell -1} & 0 \\ -U_1V_{\ell-1}^{(\ell-1)} & D_1
    \end{bmatrix} \in k[x]^{\ell r_P  \times \ell r_P},
  \end{align*}
  and $V_\ell^{(\ell)} =
  \begin{bmatrix}
   0 &   V_1^{(1)}
  \end{bmatrix} \in k[x]^{r_P \times \ell r_P}$. Then, one can check that 
  $ M_{y^{\ell}} =   V_{\ell}^{(\ell)}D_{\ell}^{-1}U_{\ell}$.

  Moreover,
  $\det D_\ell = \det(D_{\ell-1}) \det (D_1) = p_{r_P}^{\ell }$ by
  induction.  Finally, for $0 \le k < \ell$,
  \begin{align*}
    M_{y^{k}} =  V_{k}^{(\ell-1)}D_{\ell-1}^{-1}U_{\ell-1} =  V_{k}^{(\ell)}D_{\ell}^{-1}U_{\ell},
  \end{align*}
  with $V_{k}^{(\ell)} =
  \begin{bmatrix}
      V_k^{(\ell-1)} & 0_{r_P}
  \end{bmatrix}$ and this completes the proof.
\end{proof}

\begin{lemma}
  \label{lem:common_denominator_mulmat}
  Let $\ell \ge 0$. There exists a non-singular polynomial matrix
  $D \in k[x]^{r_P \times r_P}$ with $\det D = p_{r_P}^{\ell}$ such
  that for any $g \in k[x,y]$ of degree at most $\ell$ in $y$, one can
  write
  \begin{align*}
    M_g = N_g \cdot D^{-1},
  \end{align*}
  with $N_g \in k[x]^{r_P \times r_P}$ a polynomial matrix with
  $\det N_g = p_{r_P}^{\ell-\deg_y(g)}\res_y(P,g)$, where $\res_y(P,g)$ denotes the
  resultant of $P$ and $g$ w.r.t.~$y$.
\end{lemma}

\begin{proof}
  Let $g = \sum_{i=0}^\ell g_i(x)y^i \in k[x,y]$, we have
  $M_g = \sum_{i=0}^\ell g_i(x)M_{y^i}$.  Thus by
  Lemma~\ref{lem:realisation_powMy}, we obtain
  \begin{align*}
    M_g =  \left(\sum_{i=0}^\ell g_i V_i^{(\ell)} \right) D_{\ell}^{-1}U_\ell =
    V_g D_{\ell}^{-1}U_\ell,
  \end{align*}
  with $V_g \in k[x]^{r_P \times \ell r_P }$,
  $D_\ell \in k[x]^{\ell r_P \times \ell r_P}$ with
  $\det D_\ell = p_{r_P}^\ell$, and $U_\ell =
  \begin{bmatrix}
    I_{r_P} & 0
  \end{bmatrix}^t \in k[x]^{\ell r_P \times r_P }$.
  Now, since $D_\ell$ and
  $U_\ell$ are left relatively prime, by
  \cite[Thms.~8-9]{coppel1974}, there exists right relatively
  prime matrices $\tilde N \in k[x]^{\ell r_P \times r_P}$, and
  $ D \in k[x]^{r_P \times r_P}$ non-singular such that
  $D_{\ell}^{-1}U_\ell = \tilde N  D^{-1}$,
  and~$\det  D = p_{r_P}^{\ell}$. Finally, set
  $N_g = V_g \tilde N \in k[x]^{r_P \times r_P}$. We get
  $  M_g = N_g  D^{-1}$.
  Moreover, it is classical that $\det M_g = p_{r_P}^{-\deg_y(g)} \res_y(P,g)$, and thus
  $\det N_g = \det M_g \cdot \det  D =  p_{r_P}^{\ell -\deg_y(g)} \res_y(P,g)$,
  whence the result is proven.
\end{proof}

Therefore by Lemma~\ref{lem:common_denominator_mulmat}, we have a
common denominator $D$ to write all matrices of multiplication~$M_g = N_g D^{-1}$
for $g \in k[x,y]$ of degree at most $d_L$ in $y$.  As a consequence,
we have~$S_i = N_{l_i} \cdot N_{l}^{-1}$, and letting $Q = N_l$ and $N_i = N_{l_i}$
for $0 \le i < r_L$ gives the description~(\ref{eq:T_DP_description})
of $T_\DP$.
Finally,~$\det Q = p_{r_P}^{d_L- \deg_y(l(y))}\res_y(P,l(y)) $, and
bounding the degrees in $x$ in the columns of the Sylvester matrix shows that
$\deg(\res_y(P,l(y))) \le \deg_y(l(y))d_P$.  Thus,
\begin{align*}
\deg(\det Q)  \le (d_L-\deg_y(l(y)))d_P + \deg_y(l(y))d_P =
d_Ld_P,  
\end{align*}
and this completes the proof of Proposition~\ref{prop:description_T_DP}.

\section{Hermite reduction for creative telescoping}
\label{sec:hermite_red_bd}

Let $f = p/q$ be in $k(x,y) \setminus \{ 0 \}$ with $p$ and $q$
coprime in $k[x,y]$.  For simplicity, we assume that~$q$ is
square-free with respect to $y$, $\deg_y p < \deg_y q = r$ and
$\deg_x p \le \deg_x q = d$.  We also denote~$q_y = \DP_y(q)$ and
$q_x = \DP_x(q)$.  A telescoper
$L \in k[x]\left<\DP_x\right> \setminus \{ 0 \}$ for $f$ (\ie such
that $L(f)= \DP_y(h)$ for some $h \in k(x,y)$), satisfies the following
degree bound.

\begin{theorem}
  \label{thm:herm_red_bd}
  Let $f(x,y) = p/q \in k(x,y)$ with $q$ square-free with respect to
  $y$, $(d,r) = (\deg_x(q),\deg_y(q))$ and $\deg_x p \le d$,
  $\deg_y p < r$.  Then there exists a minimal telescoper for $f$ of
  order $\rho \le r$ and degree at most $\rho (d + 2rd)$.  Moreover, for
  all $m \ge \rho$ there exists a non-minimal telescoper for $f$ of
  order $m$ and degree at most $1/(m-\rho+1)\cdot(\rho d + 2mrd)$.
\end{theorem}

We retrieve the best known bound on the degree of a minimal order telescoper
for a bivariate rational
function~\cite[Thm.~25]{BoChChLi2010}, namely a degree
governed by $2d_y^2d_x$.  This bound was derived from the analysis of
the Almkvist-Zeilberger algorithm for rational
functions~\cite{AlZe1990}.
In the rest of the section, we investigate the
 Hermite
 reduction-based algorithm~\cite{BoChChLi2010} for computing a telescoper
 and prove that it can be seen as an instance of Problem~\ref{pb:lin_rel_pseudo_lin_map}.
Finally, Theorem~\ref{thm:herm_red_bd}
is proved as a consequence of Theorem~\ref{thm:degree_bound_pseudo_krylov}.
This is the first time that a degree
bound in $ O(d_y^2 d_x)$ is derived directly from the Hermite
reduction-based algorithm~(see~\cite[Lem.~18]{BoChChLi2010}).

\subsection{Reduction to Problem~\ref{pb:lin_rel_pseudo_lin_map}}
\label{sec:hermite_red_pb1}

For any rational function $g \in k(x,y)$ whose denominator is a power
of $q$, we denote by $\herm(g)$ its Hermite
reduction~\cite{hermite1872}, namely the unique polynomial
$r(x,y) \in k(x)[y]$ such that $\deg_y(r) < \deg_y(q)$ and
$g = \DP_y(h) + r/q$ for $h \in k(x,y)$.  Let $\mathcal A$ be the
vector space $k(x)[y]_{< r}$.  It is known that $\herm$ defines a
$k(x)$-linear map from $k(x)[y,q^{-1}]$ to $\mathcal A$ and
$\herm(g) = 0$ if and only if $g$ is a derivative (w.r.t.~$y$).  Moreover,
one can show~\cite{BoChChLi2010} that
$\herm(\DP_x(g)) = (\DP_x + T) ( \herm(g))$ where $T$ is the
$k(x)$-linear map:
\begin{align*}
  T \colon a \in \mathcal A \mapsto  - \herm(q_x a / q^2) \in \mathcal A.
\end{align*}
A telescoper $L$ for $f$ is such that $\herm(L( f)) = 0$, and thus
by linearity of $\herm$ it can be seen as a linear relation between
$\herm(\DP_x^i ( f))$ for $i \ge 0$.  Besides, by induction on
$i \ge 0$, we obtain~$\herm(\DP_x^i ( f)) = \theta^i(p)$ where
$p = \herm(f)$ is the numerator of $f$ and $\theta$ the pseudo-linear
map~$\DP_x + T$. Therefore, a minimal telescoper for $f$ can be
obtained as the solution of Problem~\ref{pb:lin_rel_pseudo_lin_map} on
input $(T,p)$. This is essentially the algorithm
in~\cite[Sec.~3.1]{BoChChLi2010}.

\subsection{McMillan degree}
\label{sec:McMillan_hermite}

In the basis $(1,y,\dots,y^{r-1})$, we consider $T$ as a matrix with
rational function as coefficients, in other words,~$T \in k(x)^{r \times r}$.
\begin{lemma}
  \label{lem:herm_red_XMm1Y}
  There exist matrices
  $X \in k^{r \times 2r}, M \in k[x]^{2r \times 2r}$ and
  $Y \in k[x]^{2r \times r}$, with $M$ invertible,~$\deg M \le d$, and
  $\deg Y \le d-1$, such that $T= XM^{-1}Y$.
\end{lemma}
\begin{proof}
  By analysing Hermite reduction for $q_x a /q^2$ and its associated
  linear system as
  in~\cite{horowitz1971,BoChChLi2010}, for all
  $a \in \mathcal A$ we have $q_x a /q^2 = \DP_y (A/q) + r/q$ for
  $A,r \in \mathcal A$ and $r = \herm(q_x a/q^2)$. This can be
  reformulated as
  \begin{align*}
    q_xa = q \DP_y(A) - q_y A + qr,
  \end{align*}
  just by multiplying the previous equality by $q^2$. Viewing $A$ and
  $r$ as undetermined polynomials in~$\mathcal A$,~$(A,r)$ is solution
  to the linear system
  \begin{align*}
    \begin{bmatrix}
      M_1 & M_2
    \end{bmatrix}
    \begin{bmatrix}
      A \\ r
    \end{bmatrix} = q_x a,
  \end{align*}
  where $M_1,M_2 \in k[x]^{2r \times r}$ of degree at most $r$.
  By~\cite[Lem.~6]{BoChChLi2010}, $M =
  \begin{bmatrix}
    M_1 & M_2
  \end{bmatrix}$ is invertible.
  Let~$Y$ denote the multiplication by $-q_x$, \ie $Y \colon a \mapsto -q_xa$.
  For all $a \in \mathcal A$, we have $Ta = XM^{-1}Y a$ where~$X$ is the projection according to
  the last $d_y$ coordinates, and this completes the proof since~$\deg M \le d$ and $\deg Y \le d-1$.
\end{proof}

By the same arguments as in
Section~\ref{sec:McMillan_algeqtodiffeq}, one can derive the
following bound from Lemma~\ref{lem:herm_red_XMm1Y},
\begin{align}
  \label{eq:degMM_T_hermite}
  \degMM(T) \le 2rd.
\end{align}
Finally, Theorem~\ref{thm:herm_red_bd} follows from
Theorem~\ref{thm:degree_bound_pseudo_krylov} and~(\ref{eq:degMM_T_hermite})
with  $a = p \in \mathcal A$ and $d_a \le d$.

\section{Conclusion}
\label{sec:conclusion}

Theorem~\ref{thm:degree_bound_pseudo_krylov} provides a unified approach
to establish tight
degree bounds in output of several algorithmic problems
for D-finite functions and P-recursive sequences.
In future work, we aim at exploiting the structure we enlightened in this work
for the design of new efficient algorithms. We notably intend to propose algorithms
solving Problem~\ref{pb:lin_rel_pseudo_lin_map} whose runtime is sensitive to the
size of the operator we compute.
We seek for a way of taking advantage of a \emph{small} realisation of the matrix~$T$
for solving pseudo-Krylov systems while relying on recent breakthroughs of polynomial
linear algebra.

\section*{Acknowledgement}
\label{sec:ack}

The author thanks Alin Bostan, Bruno Salvy and Gilles Villard for continuous support and
fruitful discussions and comments.

	
\bibliographystyle{abbrv} \bibliography{biblio}

\end{document}